\documentclass[12pt]{article}

\usepackage{graphicx}
\usepackage{subcaption}
\usepackage{epsfig}
\usepackage{epstopdf}
\usepackage{xcolor}

\usepackage[T1]{fontenc}

\DeclareSymbolFont{myletters}{OML}{ztmcm}{m}{it}
\DeclareMathSymbol{\uplambda}{\mathord}{myletters}{"15}
\DeclareMathSymbol{\upxi}{\mathord}{myletters}{"18}

\usepackage{amsfonts}

\usepackage{lmodern} 
\usepackage{amsmath}                                                    
\usepackage{amsthm}                                                     
\usepackage{amssymb}
\usepackage{multirow}
\numberwithin{equation}{section} 
\newcommand{\newc}{\newcommand}
\newc{\be}{\begin{equation}}
	\newc{\ee}{\end{equation}}
\newc{\bg}{\begin{gathered}}
	\newc{\eg}{\end{gathered}}
\newc{\tref}[1]{Table \ref{#1}}
\newc{\eref}[1]{Equation \eqref{#1}}
\newc{\su}[1]{$SU(#1)$}
\newc{\bm}[1]{\mathbf{#1}}
\newc{\fref}[1]{Figure \ref{#1}}

\newc{\ra}{\rightarrow}
\newc{\lra}{\leftrightarrow}
\newc{\ov}{\overline}
\newc{\ba}{\begin{eqnarray}}
	\newc{\ea}{\end{eqnarray}}
\newc{\mf}{\mathsf}

\usepackage{hyperref}

\begin{document}
	\begin{titlepage}
	\thispagestyle{empty}
		
		\vspace*{0.6cm}

		\begin{center}
			{
				\bf\large
				Gravitino Dark matter, Non Thermal Leptogenesis and Low reheating temperature in No-scale Higgs Inflation. }
			\\[12mm]
			Waqas Ahmed$^{a}$ \footnote{E-mail: \texttt{waqasmit@nankai.edu.cn\qquad ,\quad waqasmit@hbpu.edu.cn}},
				Athanasios Karozas$^{b}$ \footnote{E-mail: \texttt{akarozas@uoi.gr}} and
				George K. Leontaris$^{b}$
				\footnote{E-mail: \texttt{leonta@uoi.gr}} 
		\end{center}
		
		\vspace*{0.50cm}
		\centerline{$^{a}$ \it
			School of Mathematics and Physics, Hubei Polytechnic University, }
		\centerline{ \it
			 No. 16 North Road, Guilin, Huangshi, Hubei,
			China }
			\centerline{$^{b}$ \it
				Physics Department, Theory Division, University of Ioannina,}
			\centerline{\it
				GR-45110 Ioannina, Greece }
			\vspace*{0.2cm}
		\vspace*{0.2cm}
		\vspace*{1.20cm}

		\begin{abstract}
			\noindent
 We revisit  Higgs inflation in the framework of a minimal extension of the   Standard Model gauge symmetry by a $U(1)_{B-L}$ factor. 
 Various aspects are taken into account with particular focus on the role of the supersymmetry  breaking (SUSY) scale and the cosmological 
 constraints associated with  the gravitino. The scalar potential of the model is considered in the context of no-scale supergravity 
 consisting of the F-part constructed from the K\"ahler function, the D-terms and soft SUSY contributions.  We investigate several limiting 
 cases and by varying the SUSY scale from a few TeV up to intermediate energies, for a spectral index around  $n_s\sim 0.9655$ and reheating 
 temperature  $T_r\le 10^{9}$ GeV  we find that  the value of the tensor-to-scalar ratio ranges from  $r\approx 10^{-3}$ to $10^{-2}$.   
 Furthermore it is shown  that for certain regions of the parameter space  the   gravitino   can  live  sufficiently long and as such is a 
 potential candidate for  a dark matter component. In general, the  inflationary scenario is naturally implemented and it is consistent with 
 non-thermal leptogenesis whereas the dominant decay channel of the inflaton yields right-handed neutrinos. Other  aspects of cosmology and 
 particle physics phenomenology are briefly discussed.  Finally, we investigate  the case where the inflaton  is initially relaxed in a false
  minimum and estimate its probability to  decay to the true vacuum.

		\end{abstract}
		
	\end{titlepage}

	\vfill
	\newpage

	\setcounter{footnote}{0}

\section{Introduction }

The theory of cosmological inflation has  indisputable advantages. Among other implications, it provides convincing answers to the flatness and horizon problems~\cite{starobinsky,guth,Albrecht}, and  explains  the origin of the large scale structure of the universe.

During the last few decades numerous scenarios  have been proposed to formulate a detailed microscopic mechanism for a complete theory  of inflation with 
successful predictions for all related observables.  The Cosmic Microwave Background (CMB)  observations, however, put rather  tight constraints on many inflationary theories \cite{Ade:2018gkx,Akrami:2018odb}.  Several  models such as those with  quartic and quadratic potentials fail to satisfy  the bound on the tensor to scalar ratio, (which according to Planck 2018 results is $r<0.05$ at $95\%$ confidence level~\cite{Akrami:2018odb}), hence such models are ruled out. On the other side, one of the most successful inflationary models, which is in accordance with the Planck 2018 results,  is the Starobinsky model~\cite{starobinsky} which predicts $n_{s}=1-\frac{2}{N_{0}}$ and $ r=\frac{12}{N^{2}_{0}} \sim \left(0.002-0.004\right)$, where $N_{0}$ represents the numbers of e-folds. The Starobinsky inflationary potential has been studied  extensively in the literature within various contexts, while it was shown that this model can be derived in the context of no-scale supergravity (SUGRA) \cite{ENO,CFKN,EKN,LN,Einhorn:2009bh,Ferrara:2010in,Ferrara:2010yw,Ellis:2020lnc}.
The simplest version of the  Starobinsky proposal is equivalent to an inflationary model in which the scalar field couples non-minimally to gravity. A natural choice for the inflaton field in  Grand Unified Theories (GUTs) in the context of supergravity, is the Higgs field  breaking  the GUT symmetry~\cite{Pallis:2011gr, Ellis:2014dxa, Ellis:2016spb, Ahmed:2018jlv, Leontaris:2016jty}.
Inflation with Standard Model (SM) like Higgs boson in the
Minimal Supersymmetric Standard Model (MSSM) has also been proposed~\cite{Einhorn:2009bh,Ferrara:2010yw,Lee:2010hj}.
Also, in~\cite{Bezrukov:2008ej,Okada:2010jf}  
non-supersymmetric models with non-minimal Higgs inflation is discussed.  Several other ideas including~\cite{Liddle:2000cg} chaotic,
hybrid and  hilltop inflation have been proposed and investigated in detail. Some of these scenarios can also be realized 
in a string theory framework where the role of the inflaton field is played by some modulus.

Hybrid Inflation in particular,  is one of the most promising models of inflation, and can be naturally realized within the context of  supergravity theories. This scenario  is based on the inclusion of two scalar fields~\cite{Linde:1993cn}, with the first one realizing the slow-roll inflation and the second one, dubbed the ``waterfall'' field, triggering the end of inflationary epoch. While in the standard  hybrid inflationary scenario~\cite{Dvali:1994ms,Copeland:1994vg,Linde:1997sj} in  the supersymmetric GUTs, the symmetry is broken at the end of inflation, in the case of shifted \cite{Jeannerot:2000sv} and  smooth inflation~\cite{Lazarides:1995vr}, the symmetry breaking occurs during inflation and thus,   magnetic monopoles and other topological defects are inflated away.

On the side of particle physics, the experimental data collected at the Large Hadron Collider (LHC) so far, put strong constraints on  the conventional SUSY scenarios with superpartner masses around the electroweak scale. The so far null SUSY experimental results have triggered significant interest in alternative SUSY scale scenarios, such as high-scale SUSY and split SUSY among others 	\cite{ArkaniHamed:2004fb, Giudice:2004tc}. Taking into account the possible connection between inflationary dynamics and SUSY scale phenomenology it will be interesting to study the effects among   various SUSY scale scenarios with a successful model of inflation.

In this paper we study inflation in a  supergravity context where a no-scale K\"{a}hler potential is assumed. We consider the framework of the  SM  gauge symmetry augmented by a $ U(1)_{B-L}$ factor \cite{Buchmuller:2012wn, Buchmuller:2013dja, Ahmed:2020lua}. 
In this background, after employing the mechanism
proposed in~\cite{Dvali:1997uq, King:1997ia} for dynamically generating a $\mu$-term, we investigate the implementation of cosmic inflation and its interplay with the issues of SUSY scale, non-thermal leptogenesis and gravitino dark matter. We show that for TeV and Split SUSY scenarios the tensor-to-scalar ratio remains as low as $r\approx {\cal O}(10^{-3})$, while for a High scale SUSY scenario $r$ receives higher values of the order $r\sim \mathcal{O}(10^{-2})$ with the spectral index fixed at its central value $n_{s}=0.9655$ and reheating temperature $T_{r}< 5\times 10^{9}$ GeV. We also discuss the process of non-thermal leptogenesis and show that the dominant inflaton decay channel yields right handed neutrinos. For gravitino cosmology, we explore three possibilities: stable, unstable long-lived and unstable short-lived gravitino.  For the various cases we investigate the reheating and cosmological gravitino constraints and we find  a consistent inflationary scenario, that gives rather concrete predictions regarding supersymmetric dark matter and LHC phenomenology. 
We complete our analysis with an investigation of  certain  regions of the parameters which  yield a 
potential containing  a false minimum. We examine  possible scenaria of quantum tunneling effects 
and compute  the decay probability of such vacua    to the true vacuum.

The layout of the paper is as follows. In Sec.~2  we describe the basic features of the model including the superfields, their charge assignments, and the superpotential constrained by a
$U(1)$ R symmetry. The inflationary setup is described  in Sec.~3.  
The numerical analysis is presented in Sec.~4 containing the prospects of observing primordial gravity waves and non-thermal leptogenesis. In sections~5 and 6 we discuss the  gravitino cosmology and quantum tunneling. Our conclusions are summarized in Sec~7.

\begin{table}[!t]
	\begin{center}
		\begin{tabular}{c|c|c|c|c}\hline\hline
			{ \sc Superfields}&{\sc Representations}&\multicolumn{3}{c}{\sc
				Global Symmetries}\\\cline{3-5}	%
			&{ \sc under $G_{B-L}$ } & {\hspace*{0.3cm} $R$ \hspace*{0.3cm} }
			& {\hspace*{0.3cm}$B$\hspace*{0.3cm}} &{$ L$} \\\hline
			\multicolumn{5}{c}{\sc Matter Fields}\\\hline
			{$e^c_i$} &{$({\bf 1, 1}, 1, 1)$}& $1$&$0$ & $-1$ \\
			{$N^c_i$} &{$({\bf 1, 1}, 0, 1)$}& $1$ &$0$ & $-1$
			\\
			{$L_i$} & {$({\bf 1, 2}, -1/2, -1)$} &$1$&{$0$}&{$1$}
			\\
			{$u^c_i$} &{$({\bf 3, 1}, -2/3, -1/3)$}& $1$  &$-1/3$& $0$
			\\
			{$d^c_i$} &{$({\bf 3, 1}, 1/3, -1/3)$}& $1$ &$-1/3$& $0$
			\\
			{$Q_i$} & {$({\bf \bar 3, 2}, 1/6 ,1/3)$} &$1$ &$1/3$&{$0$}
			\\ \hline
			\multicolumn{5}{c}{\sc Higgs Fields}\\\hline
			
			{$ H_{d} $}&$({\bf 1, 2}, -1/2, 0)$& {$0$}&{$0$}&{$0$}\\
			
			{$ H_{u} $} &{$({\bf 1, 2}, 1/2, 0)$}& {$0$} & {$0$}&{$0$}\\
			\hline
			{$S$} & {$({\bf 1, 1}, 0, 0)$}&$2$ &$0$&$0$  \\
			{$\ov{H}$} &{$({\bf 1, 1}, 0, 2)$}&{$0$} & {$0$}&{$-2$}\\
			{$H$}&$({\bf 1, 1}, 0,-2)$&{$0$}&{$0$}&{$2$}\\
			\hline\hline
		\end{tabular}
	\end{center}
	\caption[]{\sl \small Superfield content of the model,  the corresponding representations under the local gauge symmetry $U_{B-L}$ and the properties with respect to the extra global symmetries.}\label{tab:themodel}
\end{table}
\renewcommand{\arraystretch}{1.}

\section{Description of the Model}

In this section we will present  the basic features regarding  the gauge symmetry and the spectrum of 
the effective model in which the inflationary scenario will be implemented.  The gauge symmetry 
is the  Standard Model gauge group   extended by a $U_{B-L}$ abelian gauge symmetry
\begin{equation}\label{eq:GBL}
G_{B-L}=SU(3)_{C}\times{SU(2)_L}\times U(1)_{Y}\times{U(1)_{B-L}}~\cdot
\end{equation}

\noindent  
The particle content of the model contains the MSSM matter and Higgs representations,  three singlets accommodating the right-handed neutrinos $N_{i}^{c}$,  a neutral scalar singlet $S$ and a pair of  Higgs singlets $H$, $\ov{H}$. These are  listed in Table \ref{tab:themodel},
where in addition to the
transformation properties under~(\ref{eq:GBL}), the `charges' under
the global symmetries $R,B,L$ also are shown. 

When the $H$, $\ov{H}$ singlet Higgs fields acquire vacuum expectation values (VEVs),  they break the $U(1)_{B-L}$ symmetry and, at the same time, they  provide Majorana masses to the right handed neutrinos.
For  the field content and the corresponding charge assignments given in Table \ref{tab:themodel}, the  renormalizable superpotential of the model is
\begin{equation}\label{wscalar1}
\begin{split}
W & = y_u {H_{u}}{q}{u}^c + y_d {H_{d}} {q}{d}^c +y_{e}{H_{d}}{L}{e}^c + y_{\nu}{H_{u}}{L}{N}^c\\
&+ \kappa S \left(\ov{H}H-M^{2}\right)+\lambda SH_{u}H_{d}+\beta H N^{c}N^{c}\; .
\end{split}
\end{equation}

\noindent The first line contains  the Yukawa sector providing Dirac masses to the up and down quarks, charged leptons and neutrinos. The corresponding Yukawa couplings are denoted  with $y_u$, $y_d$, $y_e$ and $y_{\nu}$ and the family indices  are generally suppressed for simplicity. The  tree-level terms in the second line involve the extra fields beyond the MSSM spectrum. The first term describes the standard  supersymmetric Hybrid Inflation model with $M$ being a GUT scale mass parameter and $\kappa$ a coupling constant coefficient. The second term, generates dynamically  a $\mu$-term for the model when the singlet scalar $S$ receives a non-zero VEV. The $H$ and $\ov{H}$ fields carry non-zero $B-L$ charges and their VEV's spontaneously break the $U(1)_{B-L}$ symmetry. Furthermore, $H$ is responsible for generating a Majorana mass for the right-handed neutrinos through the last term in \eqref{wscalar1}. By virtue  of  the extra global symmetries,
the model is protected from dangerous proton decay operators and  R-parity violating terms. For the same reason, bilinear terms of form $H_{u}H_{d}$ and $H\ov{H}$ are absent too. 
After this short description  of the salient features of the model,  in the next section we proceed with the inflationary dynamics.

\section{Inflation potential}

We will compute the effective scalar potential considering
contributions from the F- and D-sector, as well as soft supersymmetry breaking terms. 
The  superpotential terms  relevant for inflation are 
 \begin{equation}\label{wscalar}
 W\supset  \kappa S \left(\ov{H}H-M^{2}\right)+\lambda
 SH_{u}H_{d}+\beta H N^{c}N^{c}\; .
  \end{equation}
\noindent  We consider a  no-scale structure K\"ahler potential which, after including contributions of the relevant fields  in the present model,
  takes the form (from now on we set the reduced Planck mass to unity, $M_{Pl}=1$):
\begin{eqnarray}\label{kahler1}
\begin{split}
K &=-3 \log \left[T + T^{\ast}- \frac{1}{3}\left(H H^{\ast} + \bar{H} \bar{H}^{\ast} +H_u H_{u}^{\ast}+H_d H_{d}^{\ast}+S^{\dagger}S+ {N^c}^{\dagger}N^{c}\right)\right.\\
&\left.+\frac{\xi}{3}\left(H \bar{H} + H^{\ast} \bar{H}^{\ast}\right) +\frac{\zeta}{3}\left(H_{u} H_{d} + H_{u}^{\ast} H_{d}^{\ast}\right)\right],
\end{split}
\end{eqnarray}
\noindent where $T$, $T^{*}$ are K\"ahler  moduli fields and $\xi$, $\zeta$ are dimensionless parameters. For later convenience  we define
 \begin{eqnarray}
 \begin{split}
 \Delta &= \left[T + T^{\ast}- \frac{1}{3}\left(H H^{\ast} + \bar{H} \bar{H}^{\ast} +H_u H_{u}^{\ast}+H_d H_{d}^{\ast}+S^{\dagger}S+ {N^c}^{\dagger}N^{c}\right)\right.\\
 &\left.+\frac{\xi}{3}\left(H \bar{H} + H^{\ast} \bar{H}^{\ast}\right) +\frac{\zeta}{3}\left(H_{u} H_{d} + H_{u}^{\ast} H_{d}^{\ast}\right)\right]~,
 \end{split}
 \end{eqnarray} 
hence,   equation~(\ref{kahler1})  is simply written as $\text{K}=-3\log\Delta$. Furthermore, we introduce 
the standard definition  of the K\"ahler  function
\[ G= K+\log|W|^2\equiv  K+\log W+\log W^*.
\]
Then, the F-term potential is given by  
\ba \label{VGK}
V_{F}=e^G\left(G_iG_{i j^*}^{-1}G_{j^*}-3\right),
\ea 
where  $G_i (G_{j^*})$ are the derivatives with respect to the various scalar fields appearing \eqref{kahler1}. 
Using the above ingredients the F-term potential~(\ref{VGK})  takes the form 
\begin{eqnarray}\label{ftermpotential}
	\begin{split}
		V_{F}&=\frac{1}{\Delta^2}\left[\kappa^2 \left(M^2- H \ov{H}\right)^2 + \kappa ^2 S^2 \left(H^2+\ov{H}^2\right)+\lambda^2\left(H_{u}^2 H_{d}^2+S^2 H_{u}^2 +S^2 H_{d}^2\right)
		\right.\\
		& +\left. {N^c}^2\left(\beta^2{N^c}^2+4\beta^2\ov{H}^2+2\beta \kappa S H \right) -2\kappa\lambda H \ov{H} H_u H_{d}-2\kappa\lambda M^2 H_u H_{d}\right. ].
	\end{split}
\end{eqnarray}

When the Higgs doublets $H_u,  H_d, $ and the  various singlet fields $H, \bar H, S, N^c, $ are eliminated and only the K\"ahler moduli are present,
  the no-scale  structure of the K\"ahler potential implies that the effective F-term potential vanishes identically,  $V_{F}\equiv 0$.	
We first explore   inflation along the $H$ direction, starting with the supersymmetric  global minima of the potential.  It can be observed that the global minimum of the above potential lies at%
\be %
S^0 =H_{u}^{0}=H_d^{0}={N^{c}}^0=0~,
\ee %
 and
\begin{equation}
\begin{split}
H^{0} = 0, \quad H^{0}=\pm M,
\end{split}
\end{equation}
where $H^{0}=0$ corresponds to an extremum and $H^{0}=\pm M$  to  local minima. 

\noindent 
Next, we turn to the D-term potential.
For the fields  carrying ${B-L}$ quantum numbers, (shown in Table \ref{tab:themodel} and denoted collectively here  with $\phi_i$), the
  D-term potential is,
\begin{eqnarray}
V_{D}=\frac{1}{2} D^p_{a} D^p_{a}~,
\end{eqnarray}
\noindent where $$D_{a}^p=-g_{a}K_{,\phi_{i}}\left[t_{a}^p\right]_{i}^{j}\phi_{j}$$ for $SU(N)$ groups  and $D^{a}$ is defined as, $$D_{a}^p=-g_{a}K_{,\phi_{i}}\left[t_{a}^p\right]_{i}^{j}\phi_{j}-g_{a}q_{i} \varsigma~,$$ 
in the presence of $U(1)$ symmetries. 
Here $q_{i}$ is the charge under $U(1)$, the symbol $\varsigma$ stands for the  Fayet-Iliopoulos coupling constant and $K_{,\phi_{i}}\equiv dK/d\phi_{i}$. The $t_{a}^{p}$ are the generators of the corresponding group $G$ and $p = 1, . . . , {\rm dim}(G) $. The D-term potential can be written as,
\begin{eqnarray}
\begin{split}
V_{D}&= \frac{g_{b}^{2}}{2\Delta^2}\left[\phi_{i}^{\ast}\left(t_{b}^{p}\right)_{i}^{j}\phi_{j}-\zeta\bar{\phi}\left(t_{b}^{p}\right)_{i}^{j}\phi_{j}\right]^2 +\frac{g_{B-L}^{2}}{2\Delta^2}\left[\phi_{i}^{\ast}\left(t_{B-L}^{p}\right)_{i}^{j}\phi_{j}-\xi\bar{\phi}\left(t_{B-L}^{p}\right)_{i}^{j}\phi_{j}-q_{i} \varsigma\right]^2~,
\end{split}
\end{eqnarray}
 where $g_{b}\,( b = 1, 2, 3)$ and $g_{B-L} $ correspond to the $SU(3)_{c}$, $SU(2)_L$, $U(1)_{Y}$ and $U(1)_{B-L}$ gauge couplings  respectively. For our purposes, it suffices to work along $D$-flat directions where the D-term potential $V_{D}$ vanishes.  Restricting  to the scalar fields, we first observe that they  transform trivially under $SU(3)_c$, thus the corresponding D-term is  zero. The other  three  contributions  are
\begin{eqnarray}
\begin{split}
V_{D} &= \frac{g_{1}^{2}}{2\Delta^2}\left[\frac{1}{2}\mid H_{u}\mid^2-\frac{1}{2}\mid H_{d}\mid^2-\zeta \left(\frac{1}{2}\bar{H_{d}}H_{u}-\frac{1}{2}\bar{H_{u}}H_{d}\right) \right]\\
& +\frac{g_{2}^{8}}{2\Delta^2} \left[H_{u}^{\ast}\sigma^{p}H_{u}+ H_{d}^{\ast}\sigma^{p}H_{d}-\zeta\left(\bar{H_{d}}\sigma^{p}H_{u}+\bar{H_{u}}\sigma^{p}H_{d}\right)\right]^{2}\\
& +\frac{g_{B-L}^{2}}{2\Delta^2}\left[2\mid \bar{H}\mid^{2}-2\mid H\mid^{2}-\xi\left(2H\bar{H}-2\bar{H}H\right)+\left(q_{H}+q_{\bar{H}}\right)\varsigma\right]^2~.
\end{split}
\end{eqnarray}
Here $\sigma^{p},\, p=1,2,3$ are the $SU(2)_{L}$ generators (Pauli matrices) and  $ H_{u},H_{d}$ are the MSSM doublet Higgs 
fields which in component notation  will be  written  as
\begin{eqnarray}
\begin{split}
H_{u} &= \begin{pmatrix}
H_u^{+}\\
H_{u}^{0} 
\end{pmatrix}, \quad 
H_{d} &= \begin{pmatrix}
H_d^{0}\\
H_{d}^{-} 
\end{pmatrix}~.
\end{split}
\end{eqnarray}
Using $SU(2)_L$ transformation, it is possible to rotate the Higgs boson fields into the neutral directions $H_u^0$ and $H_d^0$. A D-flat direction can be achieved with $H_{u}^{0} = H_{d}^{0} $ and $\bar{H} =
 H $ ~\cite{Deen:2016zfr}. Finally,  we may also include contributions from explicit soft SUSY-breaking terms of the form
 \begin{eqnarray}
 \begin{split}
 V_{soft}=m^2\mid\phi_{i}\mid^{2}+\left(\kappa A_{\kappa} S H \bar{H}+\lambda  A_{\mu} SH_{u} H_{d}+\beta A_{N^{c}} \bar{H}N^{c}N^{c}-a_{s}S\kappa M^{2}+h.c\right),
 \end{split}\label{vsoft}
 \end{eqnarray}
 where $A_{\kappa}=A_{\mu}=A_{N^c}$ are the complex coefficients of the trilinear soft-SUSY-breaking terms. Along the inflationary trajectory, $S$ is zero, so the corresponding trilinear terms do not contribute. Then, for $S=0$  the total effective potential $V(\varphi)=V_{F}+V_{D}+V_{soft}$ is \\
 \begin{eqnarray}
 \begin{split}
V(\varphi)&=\frac{9\kappa^2 \left(M^2- H^{2}\right)^{2}}{\left(3(T+T^{\ast})-2(1-\xi)H^2\right)^2} +2 m^2 H^2.
\end{split}
\end{eqnarray}

\begin{figure}[t!]
	\begin{subfigure}{.5\textwidth}
		\centering
		\includegraphics[width=.95\linewidth]{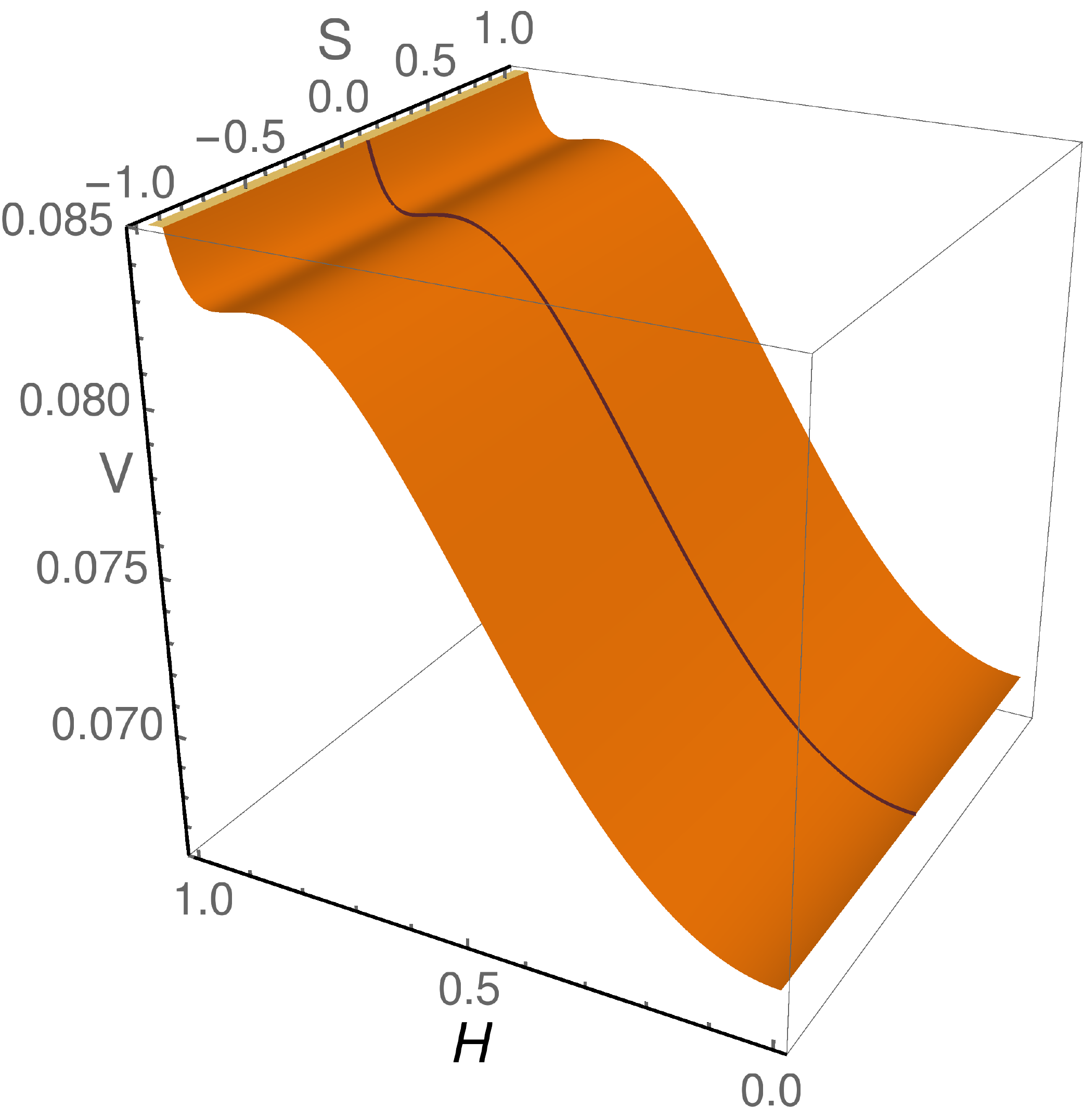}
		\label{3d1}
	\end{subfigure}%
	\begin{subfigure}{.5\textwidth}
		\centering
		\includegraphics[width=.95\linewidth]{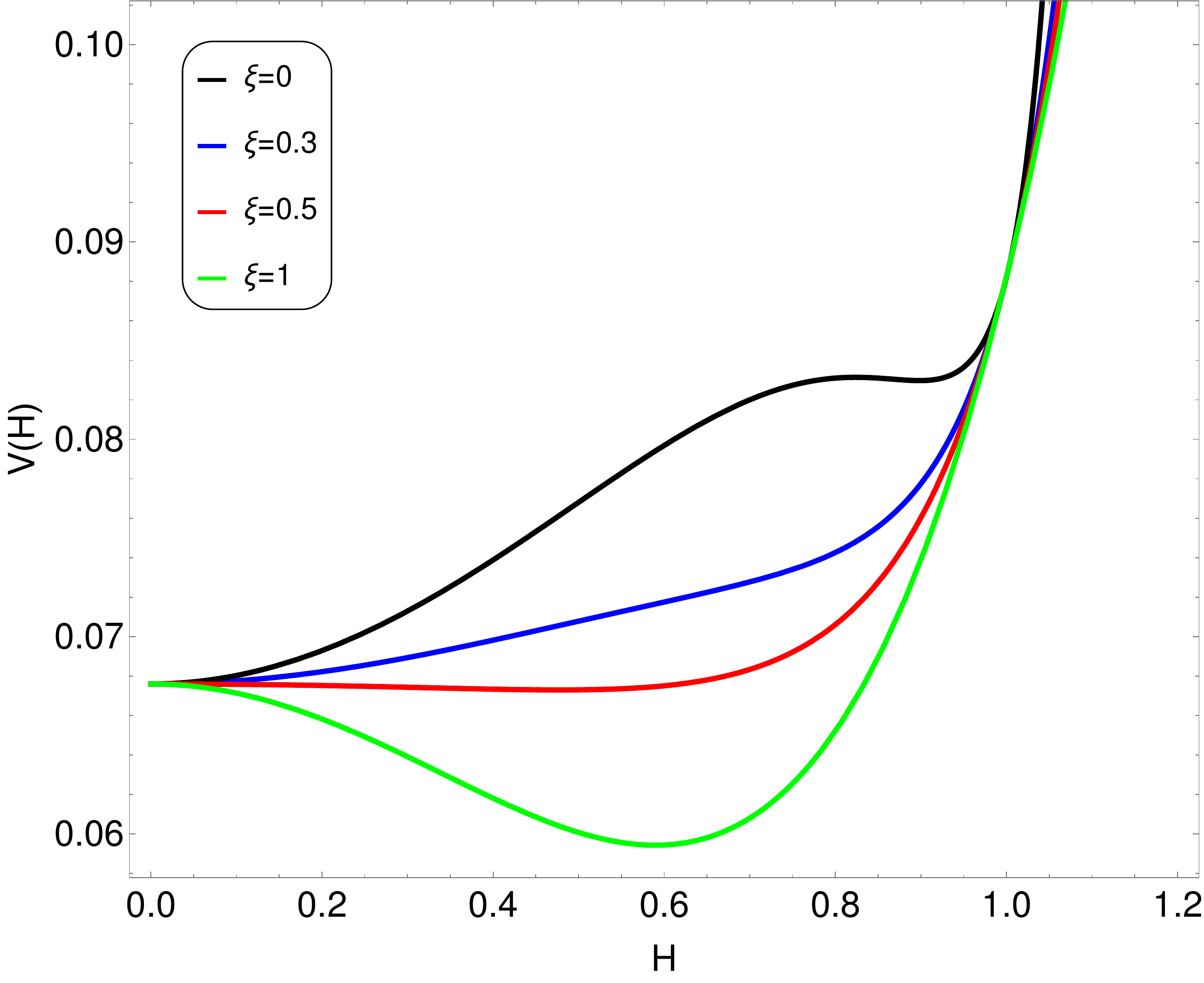}
		\label{3d2}
	\end{subfigure}	
	\caption[]{\sl \small{In the left panel we show the shape of the potential in 3-dimensions as a function of S and H. In the right panel we show the shape of the potential along the $H$ direction for different values of $\xi$. In both plots we have fixed the other parameters as: $M=1$, $m=0.2$, $\kappa=0.26$ and $T=T^{\ast}=\frac{1}{2}$. }}
	\label{3Dplots}
\end{figure}

\noindent The shape of the effective  potential is presented in Fig. \ref{3Dplots}. The left panel represents the potential as a function of $H$ and $S$ with the black curve corresponding to the $S=0$ direction. The right panel shows the potential as a function of $H$ for various values of $\xi$ and $S=0$. Along the inflationary trajectory the vacuum energy density is non-zero. The non-zero  field value during inflation also breaks $U(1)_{B-L}$ 
and the  symmetry reduces down to SM. Since the gauge symmetry is broken  during inflation, monopoles and cosmic strings are inflated away.

 We define the inflaton field as ${\varphi}=2 H$ and the modulus complex fields as $T=\left(u+i\, v\right)$, hence $T+ T^{\ast}=2u$ . Then the potential along the inflationary track is: 

\begin{equation}\label{potential_3_7}
V\left(\varphi\right) =\frac{\kappa^{2}\left(M^{2}-\frac{\varphi^{2}}{4}\right)^{2}}{\left(2u-\left(\frac{1-\xi}{6}\right)\varphi^{2}\right)^{2}}+\frac{m^2}{2} \varphi^2=\frac{V_{0}\left(1-\frac{\varphi^{2}}{4M^{2}}\right)^{2}}{\left(2u-\left(\frac{1-\xi}{6}\right)\varphi^{2}\right)^{2}}+\frac{m^2}{2}\varphi^2 \; ,
\end{equation}

\noindent where  $V_{0}=\kappa^{2}M^{4}$. However,  the inflaton field $\varphi$ is not canonically normalized since its kinetic energy terms take the  form,
\begin{equation}
\begin{split}
\mathcal{L}\left(\varphi\right)= \frac{ 2u-\frac{\varphi^2}{6}\xi\left(1-\xi\right)}{2\left(2u-\frac{1}{6}\left(1-\xi\right) \varphi^{2}\right)^{2}} \left(\partial \varphi \right)^{2} -V({\varphi}).
\end{split}
\end{equation}
We introduce a  canonically normalized field $\chi$ satisfying 
\begin{equation}\label{jphi}
\begin{split}
\left(\frac{d\chi}{d\varphi}\right)^{2} = \frac{ 2u-\frac{\varphi^2}{6}\xi\left(1-\xi\right)}{\left(2u-\frac{1}{6}\left(1-\xi\right) \varphi^{2}\right)^{2}} .
\end{split}
\end{equation}
  Integrating  (while choosing  $u=1/2$) we obtain the canonically normalized field $\chi$ as a function of $\varphi$,
\begin{equation}\label{hfield}
\chi  =\sqrt{6}\tanh^{-1}\left(\frac{\left(1 - \xi\right)\varphi}{\sqrt{6\left(1-\frac{\xi\left(1-\xi\right)\varphi^{2}}{6} \right)}}\right)
-\sqrt{\frac{6 \xi}{1-\xi}}\sinh^{-1}\left(\sqrt{\xi \left(\frac{1-\xi}{6}\right)}\varphi\right).
\end{equation}
\subsection{The $\mu$ problem and a non-zero  VEV for the $S$ field }

In section 2, we have discussed the basic features of the model 
under consideration and  introduced an $R$ symmetry to constrain 
the superpotential. In the present section we discuss the appearance
of new contributions including those  coming from supersymmetry breaking effects. 

The supersymmetric Higgs mass parameter $\mu$ (associated with the so called $\mu$ problem~\cite{Giudice:1988yz}) can be generated when 
the $S$ field acquires  a  non-zero VEV not much higher than the electroweak scale.  
Contributions from the soft SUSY breaking terms, although negligible during inflation, may generate  the required non-zero VEV, $\langle S \rangle\neq 0$~\cite{Dvali:1997uq}. 

Furthermore, in principle, we can include a constant term $W_0$ in the superpotential (\ref{wscalar1}) which gives rise to a gravitino mass \cite{Buchmuller:2000zm}-\cite{Buchmuller:2019gfy} through the formula  $m_{3/2}= e^{K/2}\langle W_0\rangle$.   The magnitude  of the constant $W_0$ is model dependent and its presence in the superpotential has important implications on the dynamics of the effective theory. 
In the present model, however, $W_0$ violates the R-symmetry of the superpotential~(\ref{wscalar1})  which has been imposed  in order to ensure linearity on $S$ and thereby  the gravitino cannot acquire a mass whithin this context. 
Nevertheless, an alternative way to generate a non-zero $m_{3/2}$ that is rather appropriate here,  is  the mechanism described in~\cite{Dvali:1997uq} (see also \cite{Pallis:2017xfo}). This relies on the soft SUSY breaking superpotential terms mentioned above which shift the VEV of the field $S$ to a non-zero value $\langle S\rangle \ne 0$.  This  is sufficient to solve the $\mu$-problem and -at the same time- generate a non-zero  mass for the gravitino.

Next, we focus on the scalar potential. 
Inserting the values  $H^{0}=\pm M, N^{0}=0$ and $H_{u}^{0}=H_{d}^{0}=0 $ the total   potential  takes the form, 
\begin{equation}\label{m32}
V_{t}(S)=\frac{2\kappa^{2}M^{2}S^{2}}{\left(2u-\frac{2M^2(1-\xi)+S^2}{3}\right)}+m_{S}^{2}S^2+\kappa A_{\kappa}M^{2}S-a_{S}\kappa M^2 +h.c.
\end{equation}
Assuming $S\ll 1$, we may expand the first term of~(\ref{m32})
in powers of $S$.  Neglecting  terms higher than  $S^2$ in the
expansion,  the potential can be expressed in terms of the gravitino mass $m_{3/2}$ as follows,
\begin{equation}\label{m322}
V_{t}(S)=\frac{2\kappa^{2}M^{2}S^{2}}{\left(2u-\frac{2 M^2(1-\xi)}{3}\right)^2}+m_{S}^{2}S^2+2 b \kappa m_{3/2} M^{2}S~,
\end{equation}
where we used $\mid a_{s}\mid+\mid A_{\kappa}\mid=2 b m_{3/2}$ while the $m_{S}^{2}$ term has been neglected since $ m_{S}^{2}\ll M^{2}$. 
Minimisation of the potential  now gives,
\begin{equation}
\sigma	\equiv 
\langle S \rangle=-\frac{ b m_{3/2}}{2\kappa }\left(2u-\frac{2M^2(1-\xi)}{3 }\right)^2~.
\end{equation} 
\noindent The second derivative with respect to $S$:
\begin{equation}
\frac{d^{2}V_t}{dS^{2}}\mid_{\sigma}=\frac{4\kappa^2 M^2}{\left(2u-\frac{2M^2(1-\xi)}{3 }\right)^2} >0~,
\end{equation}
is always positive, and the potential acquires a minimum for $\langle S\rangle \ne 0$. Hence, the $\mu$ parameter now is dynamically generated  as long as  $S$  receives a non zero VEV,
\begin{equation}
\mu=\lambda \sigma=-\frac{\lambda b m_{3/2}}{2\kappa }\left(2u-\frac{2M^2(1-\xi)}{3 u}\right)^2.
\end{equation}
Finally, for $\langle S\rangle \ne 0$  the effective potential takes  the following form,

\begin{equation}\label{potential_3_8}
\begin{split}
V\left(\varphi\right) & =\frac{\kappa^{2}\left(M^{2}-\frac{\varphi^{2}}{4}\right)^{2}+\frac{\kappa^2\sigma^2\varphi^2}{2}}{\left(2u-\left(\frac{1-\xi}{6}\right)\varphi^{2}-\frac{\sigma^2}{3}\right)^{2}}+\frac{m^2}{2} \varphi^2-\frac{ b^2 M^2 m_{3/2}^2}{2}\left(2u-\frac{2M^2(1-\xi)}{3}\right)^2 \; .
\end{split}
\end{equation}
So the Lagrangian in this case reads as,
\begin{equation}
\begin{split}
\mathcal{L}\left(\varphi\right)= \frac{ 2u-\frac{\varphi^2}{6}\xi\left(1-\xi\right)}{2\left(2u-\frac{1}{6}\left(1-\xi\right) \varphi^{2}-\frac{\sigma^2}{3}\right)^{2}} \left(\partial \varphi \right)^{2} -V({\varphi})~,
\end{split}
\end{equation}
\noindent and the canonically normalized field $\chi$ now satisfies the  equation
\begin{equation}
\begin{split}\label{cha}
\chi'\equiv\left(\frac{d\chi}{d\varphi}\right) = \sqrt{\frac{ 2u-\frac{\varphi^2}{6}\xi\left(1-\xi\right)}{\left(2u-\frac{1}{6}\left(1-\xi\right) \varphi^{2}-\frac{\sigma^2}{3}\right)^{2}}} ~\cdot 
\end{split}
\end{equation}

 \noindent We turn now to a numerical analysis and compute the various  observables
 related to infation. 

 \begin{figure}	
 	\begin{center}
 		\begin{subfigure}{.5\textwidth}
 			\centering
 			\includegraphics[width=0.9\linewidth]{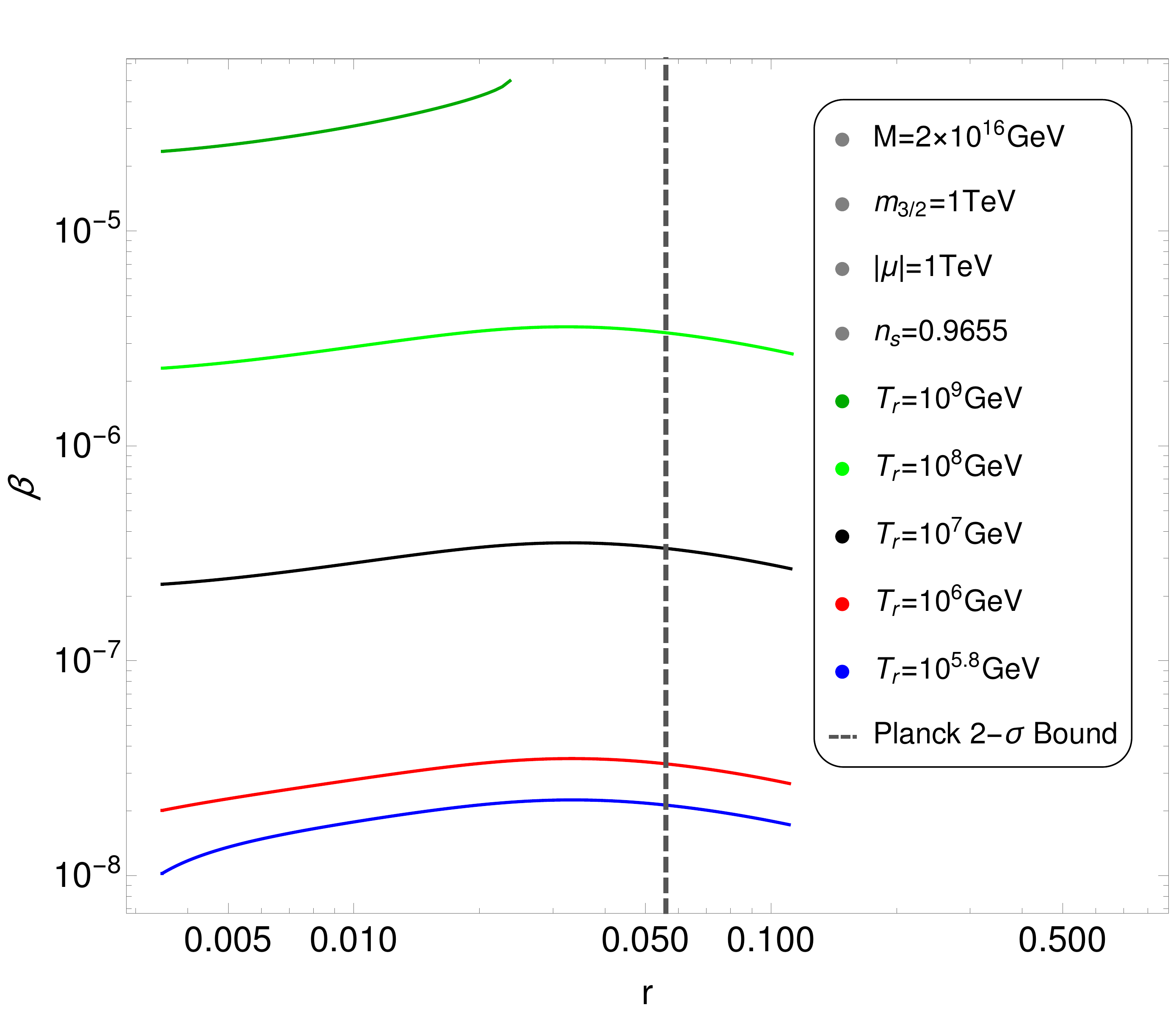}
 			\label{beta_r}
 			
 		\end{subfigure}
 	\end{center}
 	\begin{subfigure}{.5\textwidth}
 		\centering
 		\includegraphics[width=.9\linewidth]{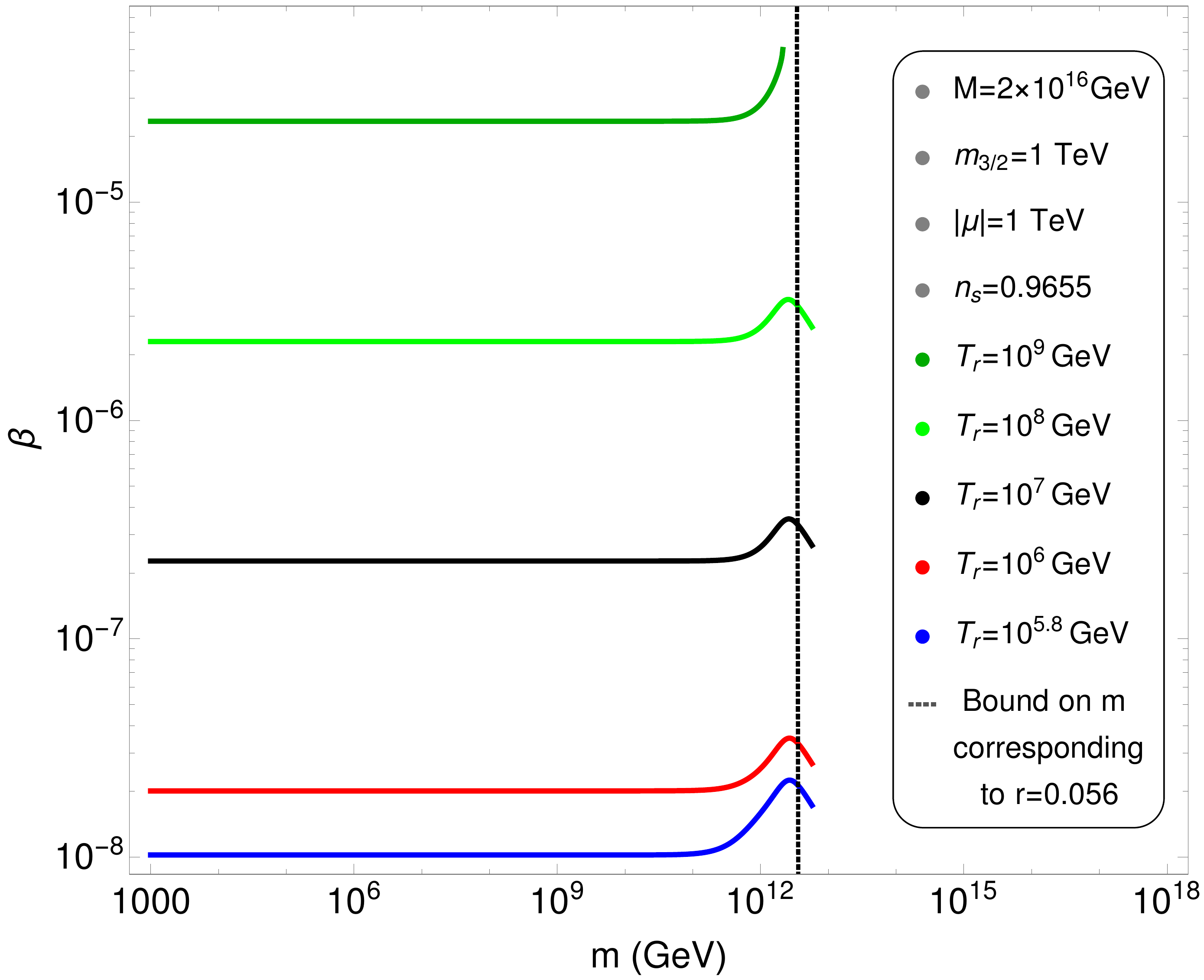}
 		\label{beta_m}
 		
 	\end{subfigure}
 	\begin{subfigure}{.5\textwidth}
 		\centering
 		\includegraphics[width=.9\linewidth]{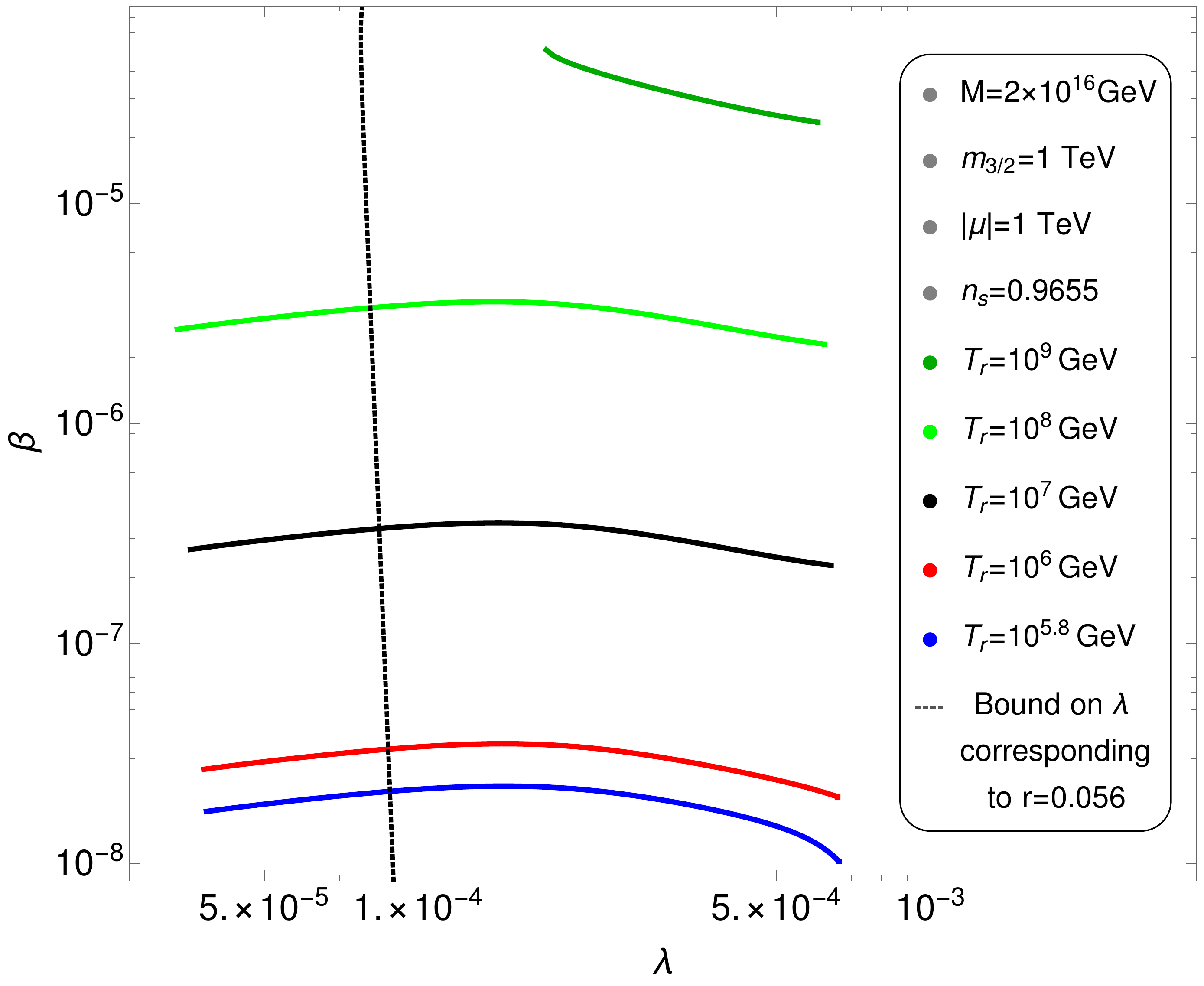}
 		\label{beta_lam}
 	\end{subfigure}\\
 	\begin{subfigure}{.5\textwidth}
 		\centering
 		\includegraphics[width=.9\linewidth]{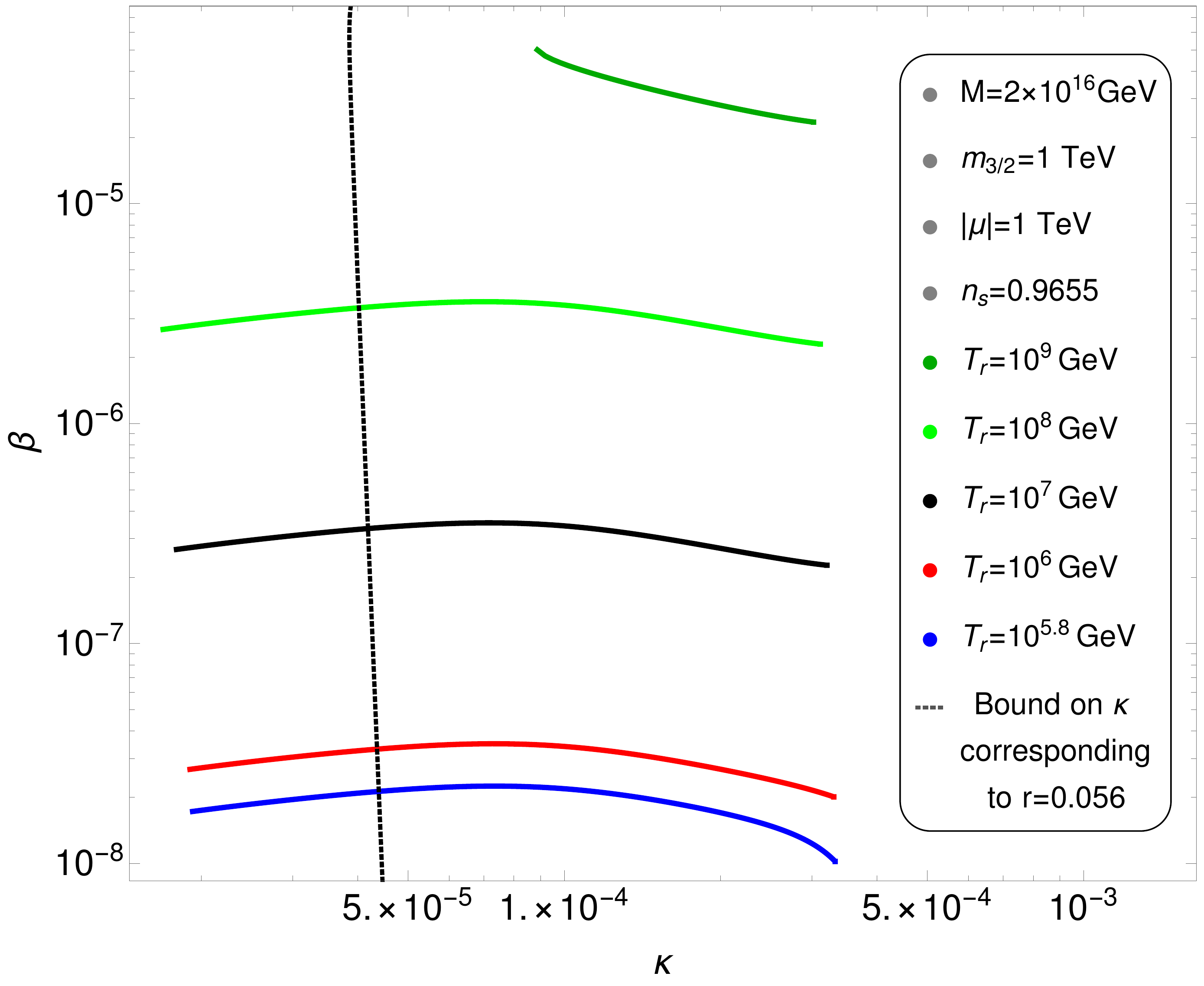}
 		\label{beta_kappa}
 	\end{subfigure}%
 	\begin{subfigure}{.5\textwidth}
 		\centering
 		\includegraphics[width=.9\linewidth]{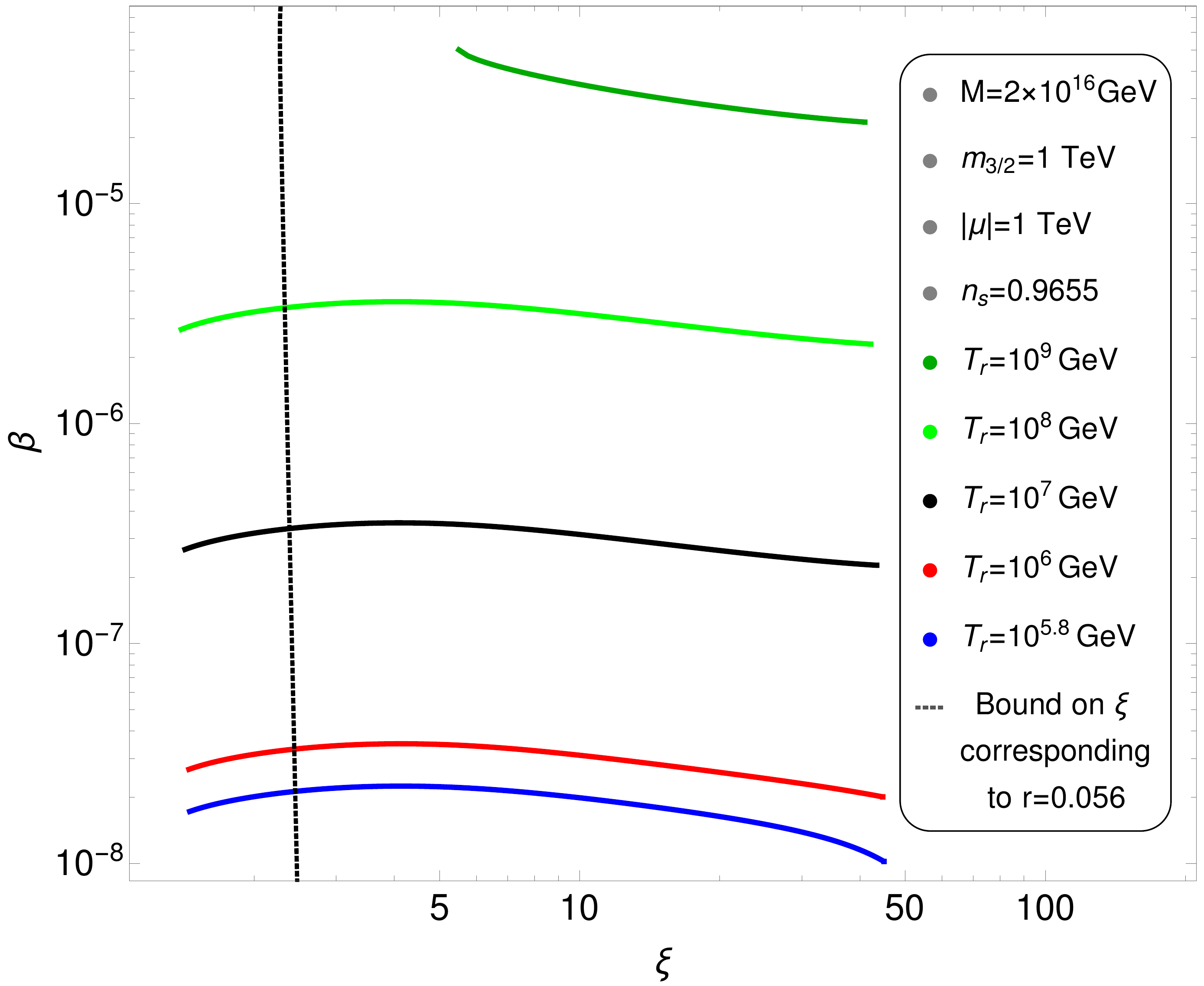}
 		\label{beta_xi}
 	\end{subfigure}
 	\caption[]{\sl \small{The plot on the top shows predictions for the tensor to scalar ratio $r$ as a function of the coupling constant  $\beta$ for various values of the reheating temperature $T_r$. Each curve corresponds to a fixed value of $T_{r}$. For all the curves we fix the GUT scale at $M=2\times{10^{16}}$ GeV, the spectral index $n_s = 0.9655$ (central value), the gravitino mass and the higgsino mass parameter at  $m_{3/2}=\mu=1$ TeV. The remaining figures on the lower  panels show the ranges for the other parameters ($m$, $\lambda$, $\kappa$, $\xi$) involved in  the analysis  with respect to the ($\beta, r$)-plot.}}
 	\label{r_plots}
 \end{figure}

\section{Analysis}

In this section we analyse the implications of the model and discuss its predictions regarding the  various
cosmological observables.  We pay particular
attention to leptogenesis, large $r$ solutions and gravitational waves.
 Before presenting numerical predictions we briefly review  the basic results of the slow-roll assumption.

\subsection{Slow-roll approximation}
 The inflationary slow-roll parameters are given by:
 \begin{equation}
 \epsilon=\dfrac{1}{2}\left(\frac{V^{\prime}\left(\varphi\right)}{V(\varphi)\chi'\left(\varphi\right)}\right)^{2}, \quad\quad \eta=\left(\frac{V^{\prime\prime}\left(\varphi\right)}{V(\varphi)\left(\chi'\left(\varphi\right)\right)^{2}}-\frac{V^{\prime}\left(\varphi\right)\chi''\left(\varphi\right)}{V(\varphi)\left(\chi'\left(\varphi\right)\right)^{3}}\right).
 \end{equation}
 The third slow-roll parameter is,
 \begin{equation}
 s^{2}=\left(\frac{V^{\prime}\left(\varphi\right)}{V(\varphi)\chi'\left(\varphi\right)}\right)\left(\frac{V^{\prime\prime\prime}\left(\varphi\right)}{V(\varphi)\left(\chi'\left(\varphi\right)\right)^{3}}-3\frac{V^{\prime\prime}\left(\varphi\right)\chi'^{\prime}\left(\varphi\right)}{V(\varphi)\left(\chi'\left(\varphi\right)\right)^{4}}+3\frac{V^{\prime}\left(\varphi\right)\left(\chi''\left(\varphi\right)\right)^{2}}{V(\varphi)\left(\chi'\left(\varphi\right)\right)^{5}}-\frac{V^{\prime}\left(\varphi\right)\chi'''\left(\varphi\right)}{V(\varphi)\left(\chi'\left(\varphi\right)\right)^{4}}\right)
 \end{equation}
 \noindent where a prime denotes a derivative with respect to $\varphi$. The slow-roll  conditions are $\epsilon\ll1$,  $\mid \eta\mid\ll1$ and $s^{2}\ll1$, while the tensor-to-scalar ratio $r$, the scalar spectral index $n_{s}$, and the running of the spectral index $\frac{dn_{s}}{d\ln k}$ are given by
 \begin{equation}
 r\simeq16 \epsilon \quad{,}\quad n_{s}\simeq 1+2\eta-6\epsilon \quad{,}\quad \frac{dn_{s}}{d\ln k}\simeq 16\epsilon\eta-24\epsilon^{2}+2s^{2}.
 \end{equation}
 The number of e-folds is given by,
 \begin{equation}
 N_{l}=\int_{\varphi_{e}}^{\varphi_{l}}\left(\frac{V\left(\varphi\right)\chi^{\prime}\left(\varphi\right)}{V^{\prime}(\varphi)}\right) d\varphi= 54+\frac{1}{3}\ln\left[\frac{T_{r}}{10^{9} {\rm GeV}}\right]+\frac{1}{3}\ln\left[\frac{V(\varphi_{l})^{1/4}}{10^{16} {\rm GeV}}\right]~.
 \end{equation}
 \noindent In the above equation,  $l$  is the comoving scale after crossing the horizon, and $\varphi_{l}$  is  the field value at  $l$. Also, $\varphi_e$ is the field at the end of inflation ends, (i.e., when  $\epsilon=1$), 
 and  $T_{r}$ is the reheating temperature which will be discussed  in the following section.
 Furthermore,  in the analysis the constraints from the amplitude of the curvature perturbation $\Delta_{R}$
 should be implemented:
 \begin{equation}
 \Delta_{R}^{2}=\frac{V\left(\varphi\right)}{24 \pi^{2} \epsilon\left(\varphi\right)}.
 \end{equation}

\subsection{Reheating Temperature and non-thermal Leptogenesis}

A successful  inflationary scenario should be followed by  thermalization  (reheating) triggered by the inflaton decay
through its couplings to SM fields and, in particular as far as the present model is concerned, the right handed neutrinos.  
This coupling,  however,  is subjected to constraints, associated with the issue of gravitino overproduction during 
thermalization~\cite{Khlopov:1984pf, Ellis:1984eq}.
 The abundance of the latter depends on the decay width of the inflaton which  is related to  the   reheating temperature.
 On the other hand, there is an upper bound on  the abundance of dark matter originating from the decay of gravitinos which 
 is converted to  the upper bound of the reheating temperature 
$T_{r}\lesssim 10^6$-$10^{11}$ GeV.  
In particular, a more precise  constraint on $T_r$ depends on the SUSY breaking mechanism and the gravitino mass $m_{3/2}$. For gravity mediated SUSY breaking models with unstable gravitinos of mass $m_{3/2}\simeq (0.1-1)$ TeV the reheating temperature  bound is  $T_r\lesssim 10^6$-$10^9$ GeV \cite{Kawasaki:1994af,Cyburt:2002uv}  while in the case of stable gravitinos raises up to $T_r\lesssim10^{10}$ GeV \cite{Fujii:2003nr}. In gauge mediated models the reheating temperature is generally more severely constrained, although $T_r\sim10^9$--$10^{10}$ GeV is possible for $m_{3/2}\simeq5$--100 GeV \cite{Gherghetta:1998tq}. Finally, the anomaly mediated symmetry breaking (AMSB) scenario may allow gravitino masses much heavier than a few TeV, thus accommodating a reheat temperature as high as $10^{11}$ GeV \cite{Gherghetta:1999sw}. In the present work we will focus on a gravity mediated SUSY breaking scenario and  in  the next section we will briefly discuss different cases for the lightest supersymmetric particle (LSP).

 The transition to the radiation epoch is controlled by the inflaton mass and its decay channels. After the end of inflation, the inflaton   starts oscillating around the minimum.  For the
 following analysis we define the canonically normalized  inflaton
 \begin{equation}
\delta\chi= \langle \chi'\rangle \delta\varphi \quad \text{with} \quad \delta\varphi=\left(\varphi-2M\right),
 \end{equation}
 where  $ \chi' $ is  defined in~ (\ref{jphi}). The non-canonical normalized fields $\varphi$  and $H$ are related to each other by $H=\varphi/2$. The magnitude of  $\langle \chi'\rangle $ at the minimum  $\varphi^{0}=2 M$  reads,
 
  \begin{equation}
   \langle\chi'\rangle\mid_{\varphi^{0}\rightarrow 2 M}= \frac{ 2u-\frac{2 M^2}{3}\xi\left(1-\xi\right)}{\left(2u-\frac{2M^{2}}{3}\left(1-\xi\right)-\frac{\sigma^2}{3} \right)^{2}}.
  \end{equation}
and the inflaton acquires a mass given by
\begin{eqnarray}
m_{\rm inf}^2=\frac{d^{2}V{(\varphi)}}{d\chi^{2}}=\left(\frac{V^{\prime\prime}\left(\varphi\right)}{(\chi')^{2}}-\frac{V^{\prime}(\varphi)\chi''}{\left(\chi'\right)^{3}}\right)_{\varphi^{0}\rightarrow 2M}.
\end{eqnarray}
Due to the superpotential terms $\beta\ov{H}N^{c}N^{c}$
and $\lambda S H_{u}H_{d}$ the possible decay channels of fields $\delta \chi,\; S$ are to 
 a pair of right handed neutrinos  and sneutrinos $(N^c, \tilde N_c)$ and  to  $H_{u}$ and $H_{d}$ respectively. The relevant Lagrangian terms are,
\begin{eqnarray}
\begin{split}
\mathcal{L}_{\delta\chi\rightarrow N^{c} N^{c}}&=-\frac{1}{2}e^{K/2}W_{,N^{c} N^{c}}N^{c} N^{c}
&\to -\alpha_{N^c}  \delta\chi N^{c} N^{c}.
\end{split}
\end{eqnarray}
with $W$ and $K$ as  defined in eqs ~(\ref{wscalar}) and ~(\ref{kahler1}), whilst $W_{,N^c, N^c}$ is the second derivative of $W$  with respect  to the field $N^c$. Finally $\alpha_{N^c}$ is the effective coupling of inflaton decay to right handed-neutrino fields which is defined as $$\alpha_{N^c}=\frac{\beta}{4\sqrt{\left(2 u-\frac{2}{3} M^2 (1-\xi )-\frac{\sigma^2}{3}\right) \left(2 u-\frac{2}{3} M^2 \xi  (1-\xi )\right)}}\; .$$
Hence,  the decay width is \cite{Pallis:2017xfo},
\begin{eqnarray}
\Gamma_{\delta\chi\rightarrow N^{c} N^{c}}=\dfrac{1}{16\pi}\alpha_{N^c}^2 m_{\rm inf}\left(1-\frac{4M_{N^c}^2}{m_{\rm inf}^2}\right)^{3/2}~,
\end{eqnarray}
where $M_{N}$ is the Majorana mass  
 $$\frac{\beta* M}{\left(2 u-\frac{2}{3} M^2 (1-\xi )-\frac{\sigma^2}{3}\right)^{3/2}}\; .$$
Similarly, for the inflaton decay to $H_{u}$ and $H_{d}$, the relevant Lagrangian is, 
\begin{eqnarray}
\begin{split}
\mathcal{L}_{\delta\chi\rightarrow H_{u} H_{d}}=-e^{K} K_{,S S^{\ast}}\mid W_{,S }\mid^2
=-\alpha_{h}  \delta\chi H_{u}H_{d}
\end{split} 
\end{eqnarray}
and the  effective coupling for inflaton decay to Higgs fields  is, $$\alpha_{h}=\frac{\kappa\lambda M\left(2u-\frac{2M^2\left(1-\xi\right)}{3}\right)}{\sqrt{\left(2 u-\frac{2}{3} M^2 \xi  (1-\xi )\right)}\left(2u-\frac{2M^2\left(1-\xi\right)+\sigma^2}{3}\right)^4}\; .$$
Thus, the decay width in this case is~\cite{Pallis:2017xfo},
\begin{eqnarray}
\Gamma_{\delta\chi\rightarrow H_{u} H_{d}}=\frac{\alpha_{h}^2}{8\pi} m_{\rm inf}.
\end{eqnarray}
 The reheating temperature $T_{r}$ , for the  $ U(1)_{B-L}$ extended MSSM spectrum is given by~\cite{Pallis:2017xfo}.
\begin{eqnarray}
T_{r}=\left(\frac{72}{5\pi^2 g^{\ast}}\right)^{1/4}\sqrt{\Gamma} \quad \text{with} \quad \Gamma=\Gamma_{\delta\chi\rightarrow N^{c} N^{c}}+\Gamma_{\delta\chi\rightarrow H_{u} H_{d}}\; .
\end{eqnarray}
 The following two conditions
 \begin{equation}\label{bound}
 \dfrac{m_{\rm inf}}{2}\geq M_{N}, \quad \quad M_{N}\geq 10\, T_{r}~,
 \end{equation}
 ensure a successful  reheating process 
with non-thermal leptogenesis. 
 These two conditions put strong bounds on the reheating temperature. 
 
 Next we present numerical predictions of the model with respect to the slow-roll parameters and the reheating constraints.

 \begin{figure}
 	\begin{subfigure}{.5\textwidth}
 		\centering
 		\includegraphics[width=.95\linewidth]{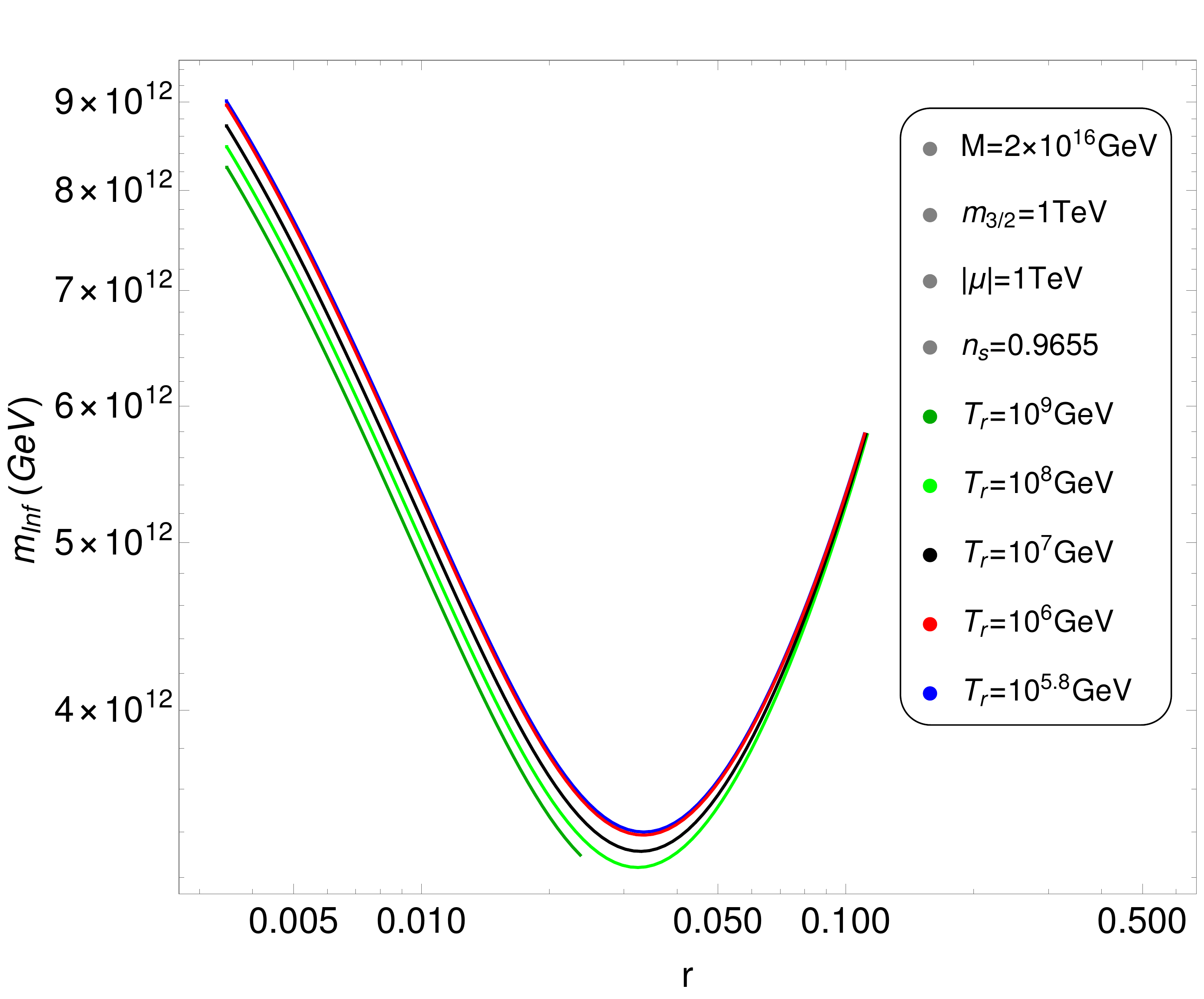}
 		\label{3d1}
 	\end{subfigure}%
 	\begin{subfigure}{.5\textwidth}
 		\centering
 		\includegraphics[width=.9\linewidth]{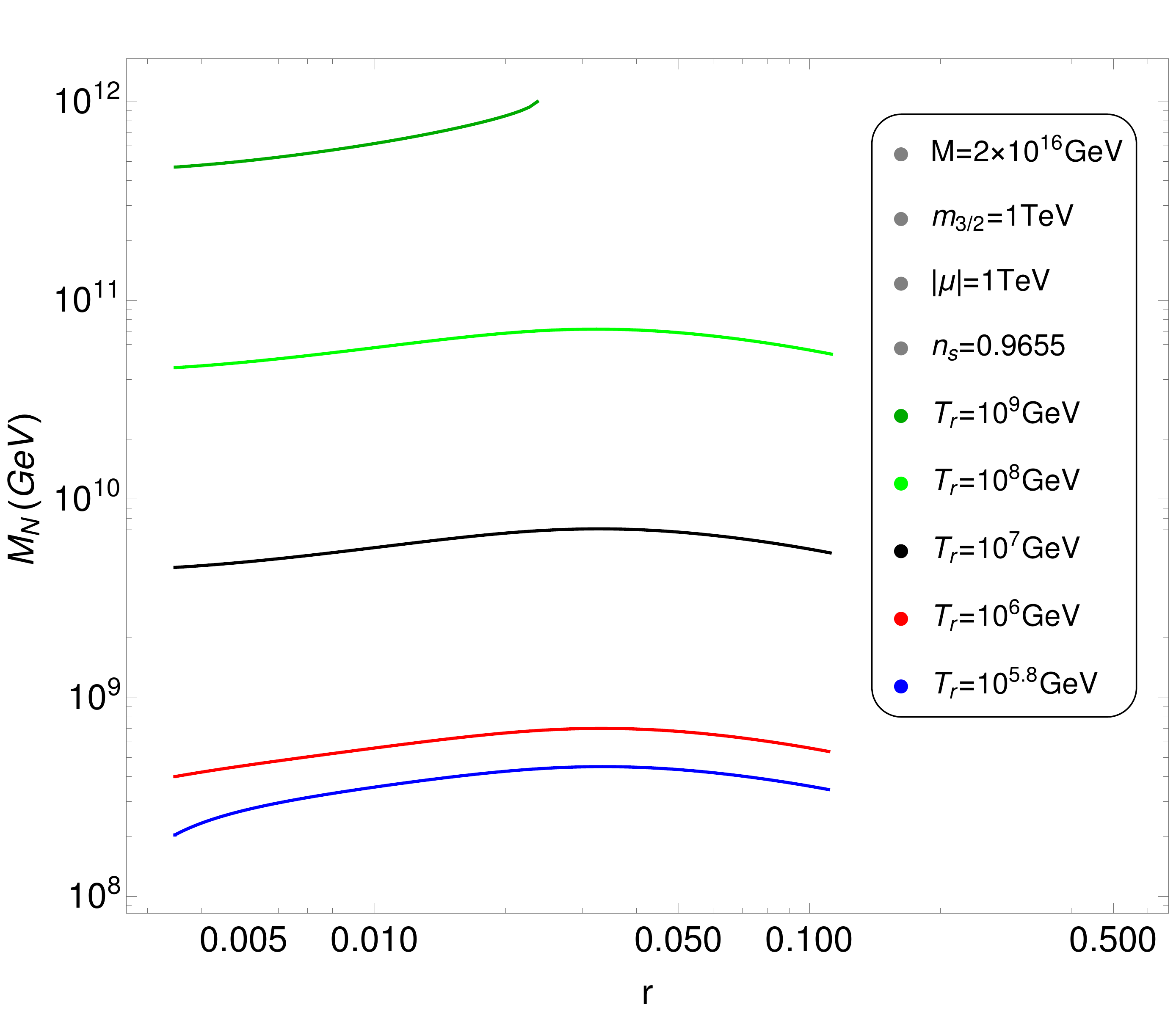}
 		\label{3d2}
 	\end{subfigure}	
 	\caption[]{\sl \small{The plot on the left shows variations of the inflaton mass $m_{\rm inf}$ as a function of the tensor to scalar ratio $r$. The right panel shows $M_{N}$ vs $r$. Each curve corresponds  to a different value for the reheating temperature $T_{r}$. }}
 	\label{minf}
 \end{figure}

 \subsection{Large $r$ solutions and observable gravity waves} 
 
 Primordial gravity waves are  associated with  the tensor-to-scalar ratio $r$ which will be measured with higher accuracy in the next-generation experiments such as PRISM, LiteBIRD, PIXIE and CORE ~\cite{Andre:2013afa,Matsumura:2013aja,Kogut:2011xw,Finelli:2016cyd}.
 Future measurements are expected to reach values as low as $r\sim 5\times 10^{-4}$.

Here, we have performed   numerical calculations to 
provide predictions for the  ratio $r$ as well as other observables. Figure \ref{r_plots} shows  ranges for the various parameters involved in the effective potential. The plot at the top panel shows  $r$ as a function of the right-handed neutrino coupling $\beta$, for various values of the reheating temperature $T_{r}$. The various curves start with  $m=1$ TeV from the left and stop when $r$ receives the value $r=0.11$. The reheating temperature varies  from $T_{r}=10^{9}$ GeV (dark green curve at the top) to $T_{r}\simeq {10^{5.8}}$ GeV (blue curve) as $\beta$ decreases. The spectral index has been fixed at $n_{s}=0.9655$ (central value) and the GUT scale mass parameter at $M=2\times10^{16}$ GeV. The remaining plots show the predictions for the  parameters $m$, $\lambda$, $\kappa$, and $\xi$. Furthermore  we consider $b=1$ and choose $\mid \mu\mid=m_{3/2}$ for each curve.   As can be observed in  Figure \ref{r_plots}   the reheating bound (\ref{bound}) is satisfied.
 
Figure \ref{minf} (left panel) shows the variations of the inflaton mass with respect to $r$. The plot shows that the mass of the inflaton  lies in the range $ 9.9\times 10^{11}\leq m_{\rm inf} \leq 9.1 \times 10^{12}$ GeV. The plot on the right shows the predictions for the right-handed heavy Majorana scale $M_{N}$ as a function of $r$. Depending on the value of the reheating temperature, $M_{N}$ varies from $10^{8}$ up to $10^{12}$ GeV, which is  heavy enough in order to realize small neutrino masses via the seesaw mechanism. In all the cases the results are in accordance with the bounds given in (\ref{bound}).

\begin{figure}
  	\begin{subfigure}{.5\textwidth}
  		\centering
  		\includegraphics[width=.95\linewidth]{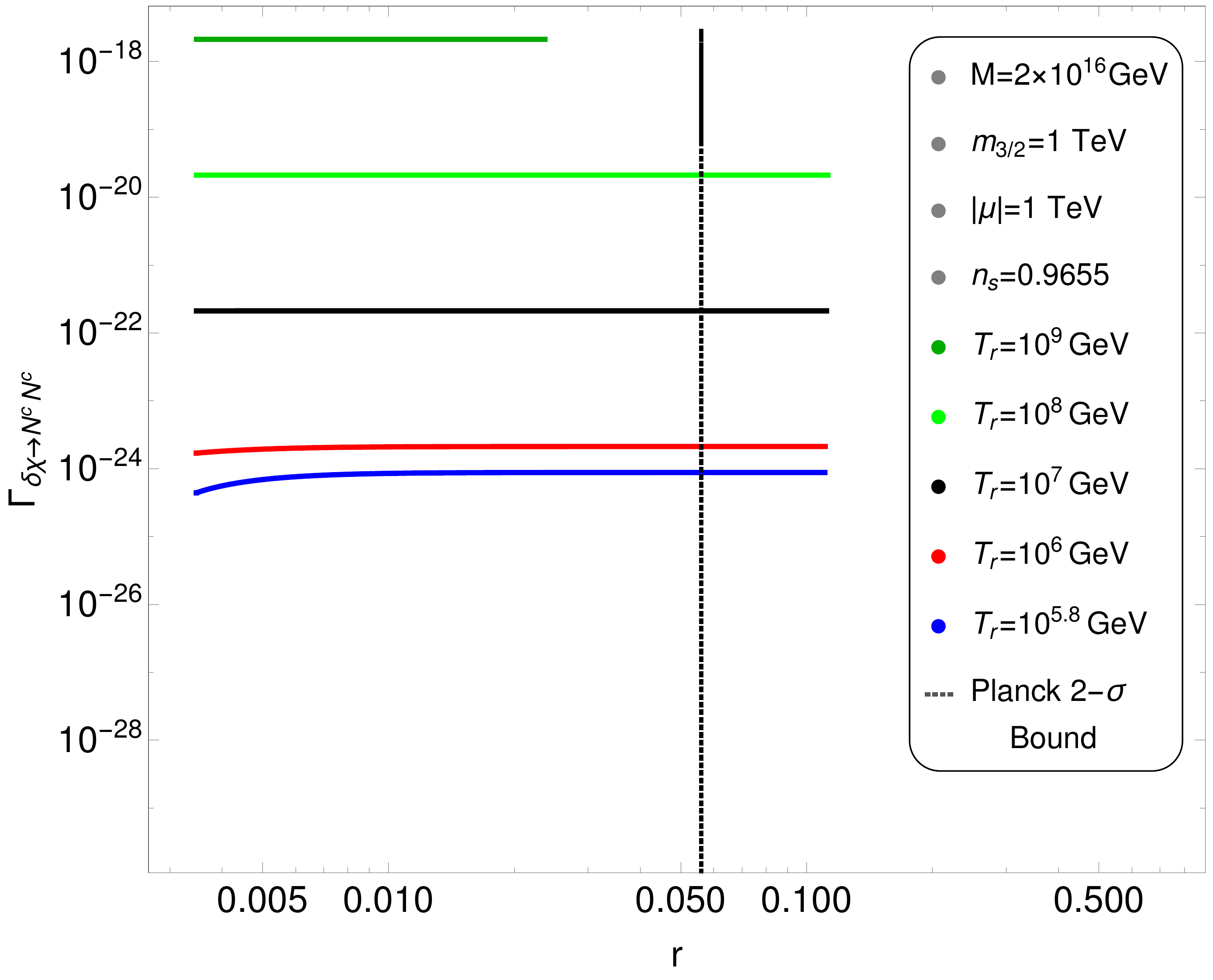}
  		\label{}
  	\end{subfigure}%
  	\begin{subfigure}{.5\textwidth}
  		\centering
  		\includegraphics[width=.95\linewidth]{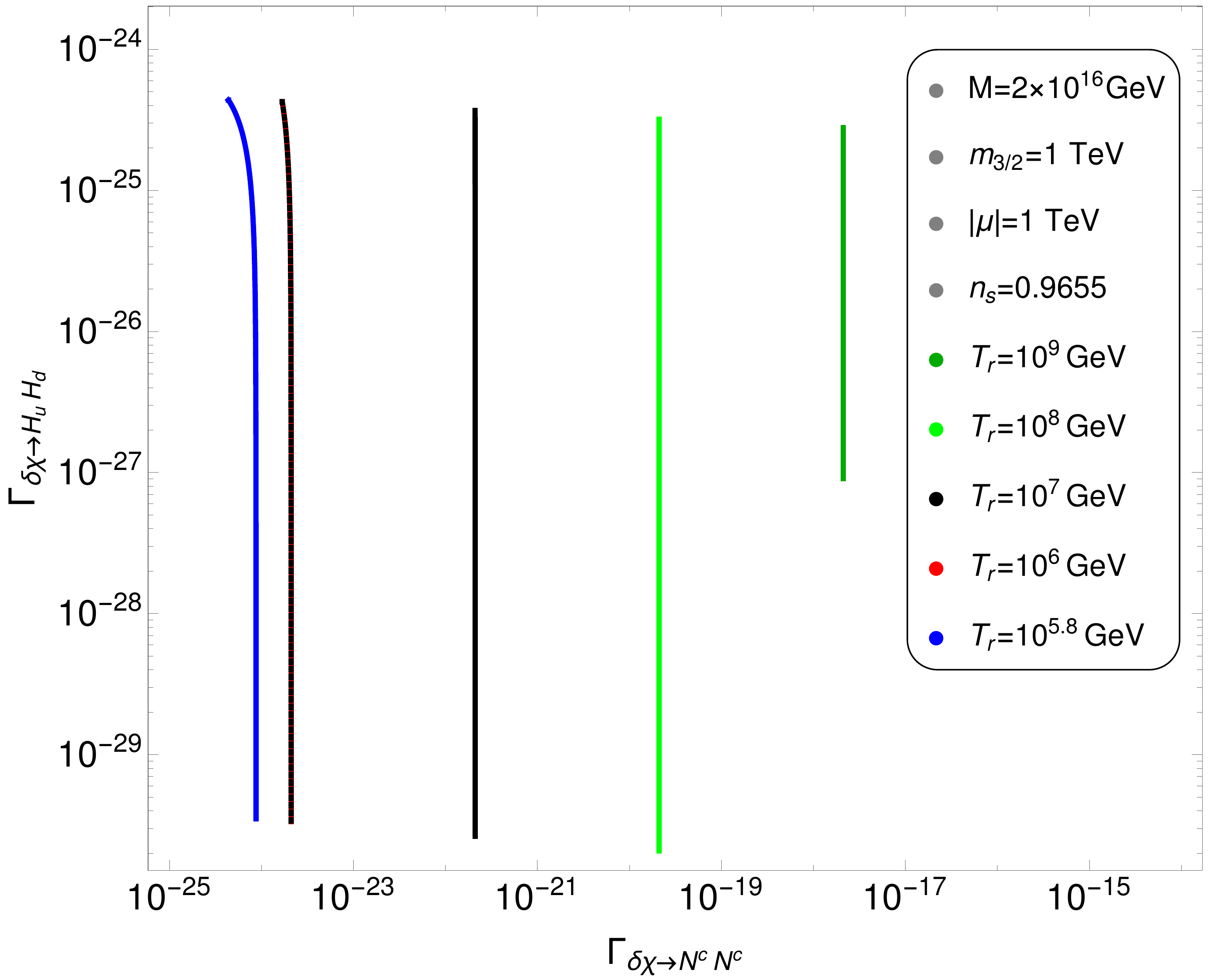}
  		\label{}
  	\end{subfigure}	
  	\caption[]{\sl \small{Left panel: plot of tensor-to-scalar ratio $r$ vs decay-width of the inflaton  to right-handed neutrino fields. Right panel:  the decay-width of the inflaton  to right-handed neutrinos vs the decay-width of the inflaton  to Higgs fields. }}
  	\label{was}
  \end{figure}

 Finally, Figure \ref{was} (left panel) shows  the  width of the inflaton decay  to Majorana neutrino with respect to $r$. In the right panel we plot the decay rate of the inflaton field to Majorana neutrinos vs the decay rate  to the Higgs fields. The dominance of the inflaton decay  to Majorana neutrinos compared to Higgs fields is reflected in this figure.
 
 \subsubsection{ Varying   the SUSY Scale; Implications}

The  results obtained show that the range of the tensor-to-scalar ratio is $r\approx [0.11-10^{-3}]$.
The Planck measurement of $r$  is up to $r= 0.056\;95\%$
 C.L. which is  shown by a gray dashed line in Figure \ref{r_plots}. In the lower (left) panel of Figure of \ref{r_plots}  the parameter $\beta$ vs $m$ is plotted. The gray line puts the cutoff for the soft breaking mass at $m_{SUSY}\leq 5\times10^{12}$ GeV. For  ${m_{SUSY}}< 10^6$ {\rm TeV}  the tensor-to-scalar ratio is around  $r\backsimeq 10^{-3}$. For high scale (split) SUSY where $m_{SUSY}\geq 10^6$ TeV we obtain a tensor-to-scalar ratio $r\backsimeq 10^{-2}$ which is in very good agreement with the measurements of $r$ in Planck and in next generation experiments.
 
  For TeV scale SUSY, Big Bang Nucleosynthesis (BBN) puts new bounds. In our model a TeV scale SUSY scenario is very consistent with our results and the observed BBN bounds. Similarly for Split SUSY and high scale SUSY scenario the reheating temperature bounded by  $\Omega_{LSP}$ which we discuss in detail in next sections.

\section{Gravitino Dark Matter}

\begin{figure}[t!]
			\centering
			\includegraphics[width=.6\linewidth]{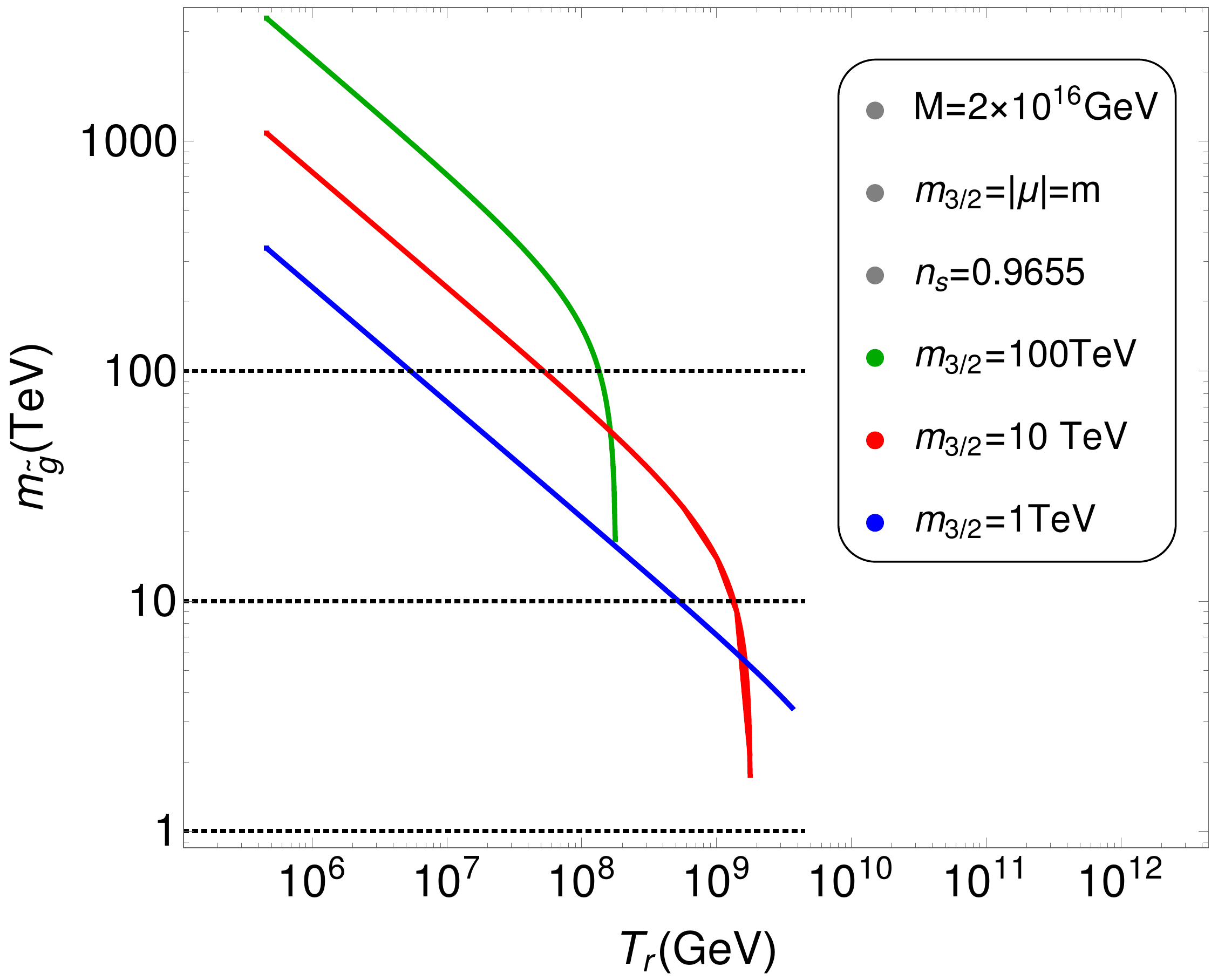}
			\caption[]{\sl \small{The case of LSP gravitino with masses $m_{3/2}=1,10, 100$ TeV. The corresponding
					gluino masses $m_{\tilde{g}}$ are shown by solid blue, red and green line with $n_{s}=0.9655$ and $M=2\times 10^{16}$ GeV.}}
			\label{mg_plot}
		\end{figure}

In this section we briefly discuss whether there are regions of the parameter space consistent with a gravitino
Dark Matter (DM) component.  According to~\cite{Lazarides:2020zof,Okada:2015vka}, one may consider the
cases of i) a stable LSP gravitino, ii) unstable long-lived gravitino with mass $m_{3/2} < 25$ TeV and,
iii) unstable short-lived gravitino with mass $m_{3/2} > 25$ TeV.
In the first case the  gravitino is a potential DM  candidate and assuming  it is  thermally produced,
its  relic density is estimated to be \cite{Bolz:2000fu}
\begin{equation}\label{omega}
\Omega_{3/2}^{2}=0.08\left(\frac{T_{r}}{10^{10} GeV }\right)\left(\frac{m_{3/2}}{1000 GeV}\right)\left(1+\frac{m_{\tilde{g}^{2}}}{3m_{3/2}^{2}}\right)~,
\end{equation}
where   $m_{\tilde{g}}$ is the gluino mass parameter and  for simplicity $m_{3/2}= m=\mid\mu\mid$ is assumed\footnote{Eq. (\ref{omega}) contains only the dominant QCD contributions for the gravitino production rate. In principle there are extra contributions descending from the electroweak sector as mentioned in \cite{Pradler:2006qh}, \cite{Rychkov:2007uq} and recently revised in \cite{Eberl:2020fml}. If we consider these type of contributions in our analysis, we estimate that (depending on gaugino universality condition) our results  will deviate $\sim{(10-15)\%}$.}. 
A stable  LSP  gravitino requires $m_{\tilde{g}}>m$ while current LHC  bounds on the gluino mass are around on $2.2$ TeV \cite{Vami:2019slp, Aaboud:2017vwy}.  Taking the  lower bound of relic abundance $\Omega^{2}_{h}=0.144$ \cite{Akrami:2018odb}, Figure \ref{mg_plot} shows the range of  $m_{\tilde{g}}$ as a function of the reheating temperature $T_{r}$ for representative values of $m_{3/2}$.

  It is seen from  Figure \ref{mg_plot} that for $m_{3/2}=1\text{TeV}$ (blue curve), the gravitino is the LSP since the kinematic condition $m_{3/2}<m_{\tilde{g}}$ is always true.  For $m_{3/2}=10$ TeV (red curve) the gravitino is the LSP in the regions with $m_{\tilde{g}}>10$ TeV. Below the $m_{\tilde{g}}=10$ TeV dotted black line the gravitino is the next lightest supersymmetric particle (NLSP). A similar description holds for the case with $m_{3/2}=100$ TeV (green curve). Therefore, we see that there are regions in the parameter space where gravitino is the LSP and as such it contributes to DM.

	 Next we consider the posibillity where gravitino is not the LSP. The decay of the gravitino occurs after the freeze-out epoch of the lightest neutralino which will play the role of LSP. The lightest neutralino has two origins; one is thermal relic and the other is non-thermal. Thermal production consists of the standard freeze-out mechanism of weakly interacting massive particles, whereas non-thermal production deals with the decay product of gravitino produced during the reheating process~\cite{Kawasaki:2004qu}.  However, since the density of the thermal relic is strongly model dependent~\footnote{For a detailed analysis with emphasis on the DM phenomenology of the model see~\cite{Ahmed:2020lua}. }, we do not take into account its effect in the calculation of the density parameter. Here we distinguish two cases of gravitino decay, either a  long-lived or a short-lived. For a long-lived gravitino with mass $m_{3/2}<25$ TeV  its lifetime  is about  $\tilde{\tau}\gtrsim 1 sec$. However, in this case we encounter the cosmological gravitino problem \cite{Khlopov:1984pf} that originates due to the fast decay of gravitino which may  affect the light nuclei abundances and thereby ruin the success of BBN theory. To avoid this problem, one has to take into account the BBN  bounds on the reheating temperature which are \cite{Kawasaki:2004qu}
		\begin{equation} \label{eqlong}
		\begin{split}
		T_{r}&  \lesssim3\times \left(10^{5}-10^{6}\right) {\rm GeV \quad for }\quad m_{3/2}= 1\; {\rm TeV }\; ,\\
		T_{r}& \lesssim 2.5\times10^{9} \;{\rm GeV } \quad {\rm for} \quad m_{3/2}=10\; {\rm TeV. }
		\end{split}
		\end{equation}

We see from Figure \ref{mg_plot} that a long lived gravitino with $m_{3/2}=1$ TeV   is not a viable scenario  because it becomes NLSP for a reheating temperature $T_{r}\geq2\times 10^{9}$ GeV. Nevertheless for $m_{3/2}\geq 10$ TeV, a long-lived gravitino scenario is viable and  consistent with the BBN bounds \eqref{eqlong} for  the reheating temperature.

		\begin{figure}[t!]
			\centering
			\includegraphics[width=.6\linewidth]{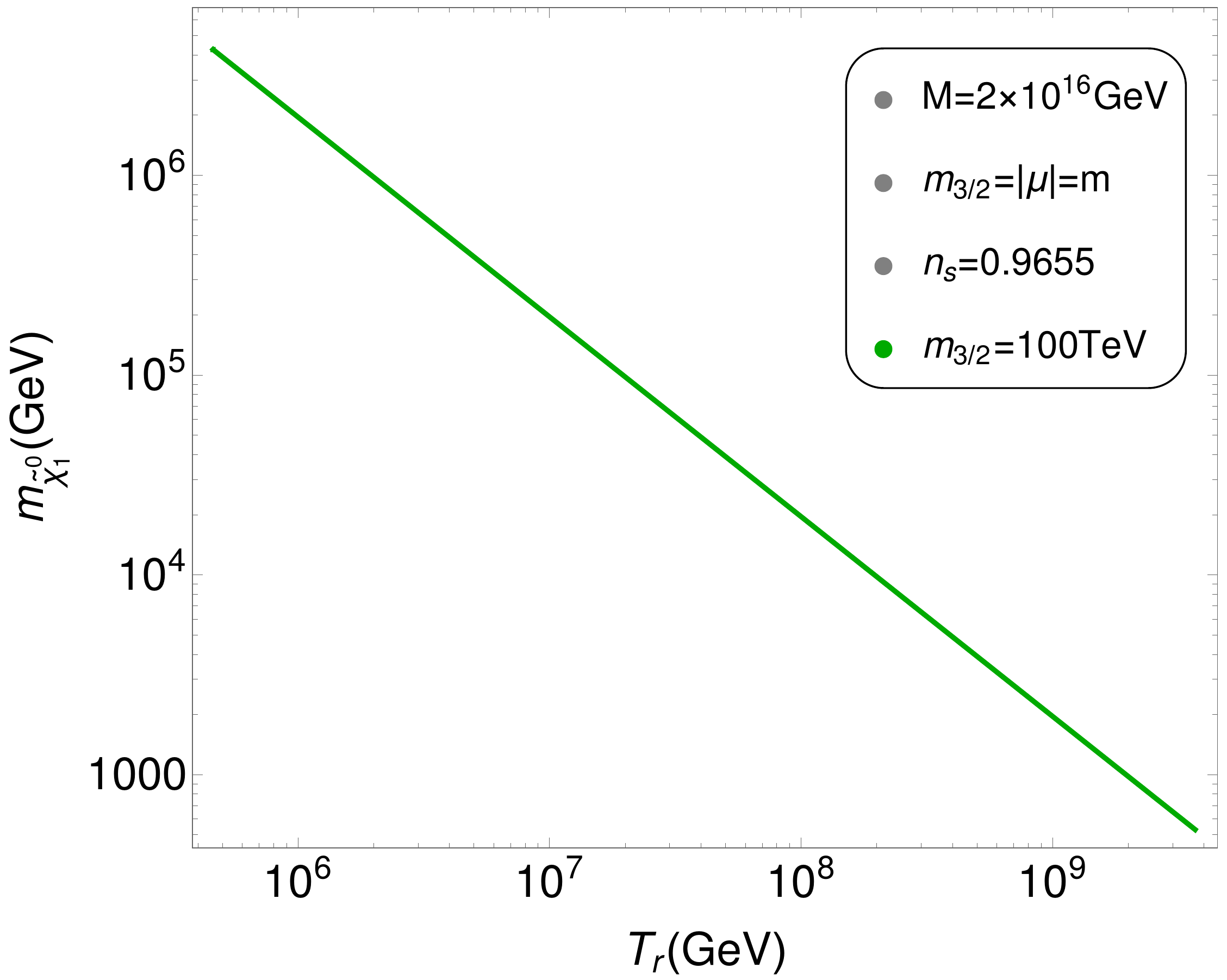}
			\caption[]{\sl\small{Short-lived gravitino with mass $m_{3/2}= 100$ TeV and the corresponding neutralino masses $ m_{\tilde{\chi}_{1}^{0}}$ with $n_{s}=0.9655$ and $M=2\times 10^{16}$ GeV.}}
			\label{mneu_plot}
		\end{figure}

In order to discuss the scenario of a short-lived gravitino (for instance with mass $m_{3/2} = 100$ TeV), we recall that 
		the gravitino decays before BBN, and as a result the BBN bounds on the reheating temperature are not effective. In this case the gravitino decays into the LSP neutralino $\tilde{\chi}_{1}^{0}$. The neutralino abundance is given by
		
		\begin{equation}\label{eqa}
		\Omega_{\tilde{\chi}_{1}^{0}}h^{2}\simeq 2.8\times10^{11}\times Y_{3/2}\left(\frac{m_{\tilde{\chi}_{1}^{0}}}{1 TeV}\right),
		\end{equation}
		where $Y_{3/2}$ is the gravitino Yield and is defined as,
		\begin{equation}\label{eqb}
		Y_{3/2}\simeq2.3\times10^{-12}\left(\frac{T_{r}}{10^{10} GeV}\right).
		\end{equation}
		
		As we know the LSP neutralino density produced by gravitino decay should not exceed the observed DM relic density. Choosing the upper bound of relic abundance $\Omega_{\tilde{\chi}_{1}^{0}}h^{2}=0.126$ and using equations (\ref{eqb}) and (\ref{eqa}), we find a relation between the reheating temperature $T_{r}$ and $m_{\tilde{\chi}_{1}^{0}}$
		which is
		\begin{eqnarray}\label{eqc}
		m_{\tilde{\chi}_{1}^{0}}\simeq19.6\left(\frac{10^{11} GeV}{T_{r}}\right)~.
		\end{eqnarray}
		For gravity mediation scenario $ m_{\tilde{\chi}_{1}^{0}}\geq18$ GeV \cite{Hooper:2002nq}, which can easily satisfied. For a gravitino with $m_{3/2}=100$ TeV, the bounds on the  LSP neutralino mass derived as in eq~(\ref{eqc}) 
		\begin{eqnarray}
		5.21\times10^{2} \leq m_{\tilde{\chi}_{1}^{0}} (\text{GeV})\leq4.25\times10^6~.
		\end{eqnarray}
		Figure \ref{mneu_plot} shows the plot between  $m_{\tilde{\chi}_{1}^{0}}$ and $T_{r}$. 
		Therefore, the short-lived gravitino scenario is also a viable possibility  in this  model.

  \section{Quantum Tunneling from False to True Vacuum }
  
  A careful consideration of the previous analysis reveals regions of the parameter space where, in addition to 
  the global minimum  $V_{min}(\varphi_0)$ (true vacuum),  a second (higher) ground state  evolves at some
  point $\varphi\ne \varphi_0$, usually called  `the false vacuum'.
   Depending on the initial conditions the inflaton may roll down the slope
 of the potential  towards the false vacuum and subsequently make the
 transition to the true minimum. Such transitions can be
 described by Euclidean instantons in a way suggested by Coleman 
 and deLucia (CdL) \cite{Coleman:1980aw}.

  In order to discuss the dynamics of the vacuum in this new context,  we rewrite the non-canonical normalized potential of \eqref{potential_3_7},  in the form
 \begin{equation}\label{newff}
  V\left(\varPhi\right) =\frac{1}{2}\mu^2\varPhi^2+\frac{\lambda}4\frac{\left(1-\varPhi^2\right)^2}{\left(1-\gamma\varPhi^2\right)^2}~,
  \end{equation}  
where we have made the following redefinitions
  \begin{equation}
  \varPhi=\frac{\varphi}{2M},\quad
  \gamma=\frac{(1-\xi)M^2}{3u}, \quad \mu=2mM, \quad \lambda=\left(\frac{\kappa M^2}{u}\right)^2~.
  \end{equation} 
 Since $m < M < 1 $, the parameters $\mu $ and  $\lambda $ are expected to be less than unity.
   
  For the special case  $\xi =1$ we have $\gamma = 0$ and the potential simplifies to
  \begin{equation}\label{p1}
  V\left(\varPhi\right) =\frac{1}{2}\mu^2\varPhi^2+\frac{\lambda}{4}\left(1-\varPhi^2\right)^2~.
  \end{equation}
  Similarly in large limits of $|\xi|$ we have $\gamma=|\xi| M^2/(3u)$. 
 The form of $V$ in eq \eqref{p1} represents a double  well potential, with the minima at a radius  $\varPhi_{0}=\sqrt{1-\frac{\mu^2}{\lambda}}$ and local maximum at $\varPhi=0$. 
In this new parametrisation the general case corresponds to $\gamma\neq0$ or $\xi\neq1$. The extrema are determined by
  \begin{equation}
  \dfrac{dV\left(\varPhi\right)}{d\varPhi}=0\Rightarrow \mu^2\varPhi\left(1 - \frac{\lambda}{\mu^2}\dfrac{\left(1-\gamma\right)\left(1-\varPhi^2\right)}{\left(1-\gamma\varPhi^2\right)^3}\right)=0\label{derv0} .
  \end{equation}
  There is an obvious extremum (in fact minimum) at $\varPhi = 0$. The parenthesis includes a cubic equation with respect to $\varPhi^2$ implying three roots for $\varPhi^2$. The real roots are either three or a single one depending on the values of the parameter $\gamma$ and the ratio, $\mu^2/\lambda=R$. The case of the matastable dS vacuum corresponds to the case of  three real roots, since  there should
  be a minimum and a maximum before $\varPhi=0$ and $\varPhi=\infty$. To simplify the cubic equation we make the redefinitions,
  
  \begin{equation}
  y=1-\gamma\varPhi^2\longrightarrow \varPhi^2=\frac{1-y}{\gamma}, \quad \text{for} \quad\gamma\neq0,\quad \text{or}\quad \xi\neq~1~.
  \end{equation}
  Then the equation \eqref{derv0} is written as
  \begin{equation}
  y^3+py+ q=0~,
  \end{equation}
  with 
  \begin{equation}
  \quad p=-\frac{1-\gamma}{\gamma R}, \quad q=-(1-\gamma)p=\frac{(1-\gamma)^2}{\gamma R}=\frac{a^2}{\gamma R}~.
  \end{equation}
  Three real roots exist if the discriminant $4p^3+27q^2<0$, implying $$\left(a/\gamma R\right)^3\left(-1+27aR\gamma/4\right)<0~.$$ Substituting $y = v \cos \theta$ and comparing with the trigonometric identity
 $ \text{ cos}(3\theta)=4\text{co}s^3(\theta)-3 \text{cos}(\theta)$,
  we find
  \begin{align*} 
  y&=2 \sqrt{\dfrac{-p}{3}} \cos \left(\frac{1}{3} \arccos \left(\frac{3q}{p} \sqrt{-\frac{3}{4p}}\right)-\frac{2n\pi}{3}\right), \quad n=0,1,2,
  \end{align*}
  or
  \begin{align*} 
  y&=2 \sqrt{\dfrac{1-\gamma}{3\gamma R}} \cos \left(\frac{1}{3} \arccos \left(-\frac{3}{2} \sqrt{3R\gamma(1-\gamma)}\right)-\frac{2n\pi}{3}\right), \quad n=0,1,2~.
  \end{align*}  
  Real roots exist when the argument of $\arccos$ is less than one,
  \begin{equation}\label{cons}
  \mid-\frac{3}{2} \sqrt{3R\gamma(1-\gamma)} \mid\leq1,
  \end{equation}
  which is just the constraint on the discriminant. Assuming $0 < R < 1$ and using \eqref{cons} we can put bounds on the various parameters of the potential. The above inequality can be simplified as,
  \begin{equation}
  \gamma^2-\gamma+\dfrac{4}{27R}\geq 0~.
  \end{equation}
  For real roots of $\gamma$ the discriminant must be greater or equal to zero which means $1-\frac{16}{27R}\geq0$ or  $R\geq \frac{16}{27}$. This constrains $R$ in the range $ 16/27\leq R<1$, or in terms of $\mu$ and $\lambda$ we have that
  \begin{equation}
   \mu^2< \lambda\leq\frac{27}{16}\mu^2 \; .
  \end{equation}
  
 The bound on $R$ also helps us to obtain a  lower bound on $\gamma$ which is $\gamma \ge \frac 12 (1-\sqrt{\frac{11}{27}})$. The upper value of $\gamma$ basically separates the potential into two regions.
  Then,   the relation \eqref{jphi} between the canonically and the non-canonically normalized field with respect to the  new parametrization  can be written as,
  \begin{equation}\label{con-non}
  \dfrac{dX}{d\varPhi}=\sqrt{\frac{1-\gamma\varPhi^2\left(1-\frac{3u}{M^2}\gamma\right)}{2u\left(1-\gamma\varPhi^2\right)^2}}\; ,
  \end{equation}
  
\noindent where $X=\chi/2M$ is the normalized field.

  We solve eq. (\ref{con-non}) in two limits. In the first case we consider a large field approximation with  
  \begin{eqnarray}
  \varPhi\gg\frac{1}{\sqrt{\gamma}}
  \end{eqnarray}
  
 \noindent and  we find that,
  \begin{eqnarray}
  \dfrac{dX}{d\varPhi}=\sqrt{\frac{\left(\frac{3u}{M^2}\gamma-1\right)}{2u\gamma}}\frac{1}{\varPhi}\quad\Longrightarrow X=\sqrt{\frac{\left(\frac{3u}{M^2}\gamma-1\right)}{2u\gamma}}\ln(\sqrt{\gamma}\varPhi)\; .
  \end{eqnarray}
  In terms of $\varPhi$, we have that
  \begin{eqnarray}
  \varPhi=\frac{1}{\sqrt{\gamma}}\,\exp\left({\sqrt{\frac{2u\gamma}{\left(\frac{3u}{M^2}\gamma-1\right)}}}\,{X}\right)\; .
  \end{eqnarray}
  The effective potential ~(\ref{newff}) in terms of the canonically normalized field $X$  can be written as,
  \begin{equation}\label{Vlfa}
  V\left(X\right) =\frac{\mu^2}{2\gamma} \,\exp\left({\sqrt{\frac{2u\gamma}{\frac{3u}{M^2}\gamma-1}}}\,{2X}\right)+\frac{\lambda\left(1-\frac{1}{\gamma} \exp\left({\sqrt{\frac{2u\gamma}{\frac{3u}{M^2}\gamma-1}}}\,{2X}\right)\right)^2}{4\left(1-\exp\left({\sqrt{\frac{2u\gamma}{\frac{3u}{M^2}\gamma-1}}}\,{2X}\right)\right)^2}~\cdot 
  \end{equation}
 
\begin{figure}[t!]
  	\centering
  	\includegraphics[width=0.49\linewidth]{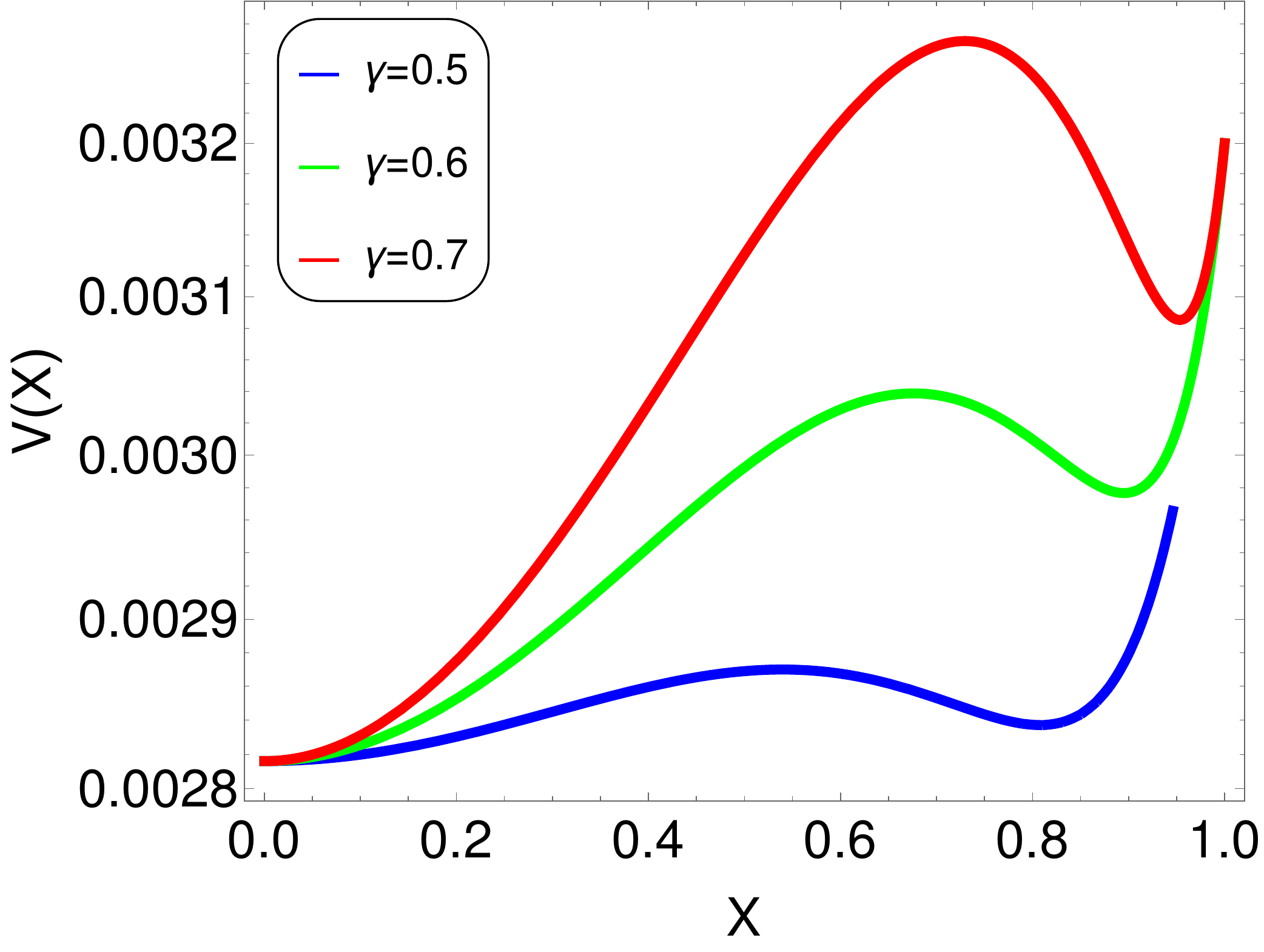}%
  	\includegraphics[width=0.49\linewidth]{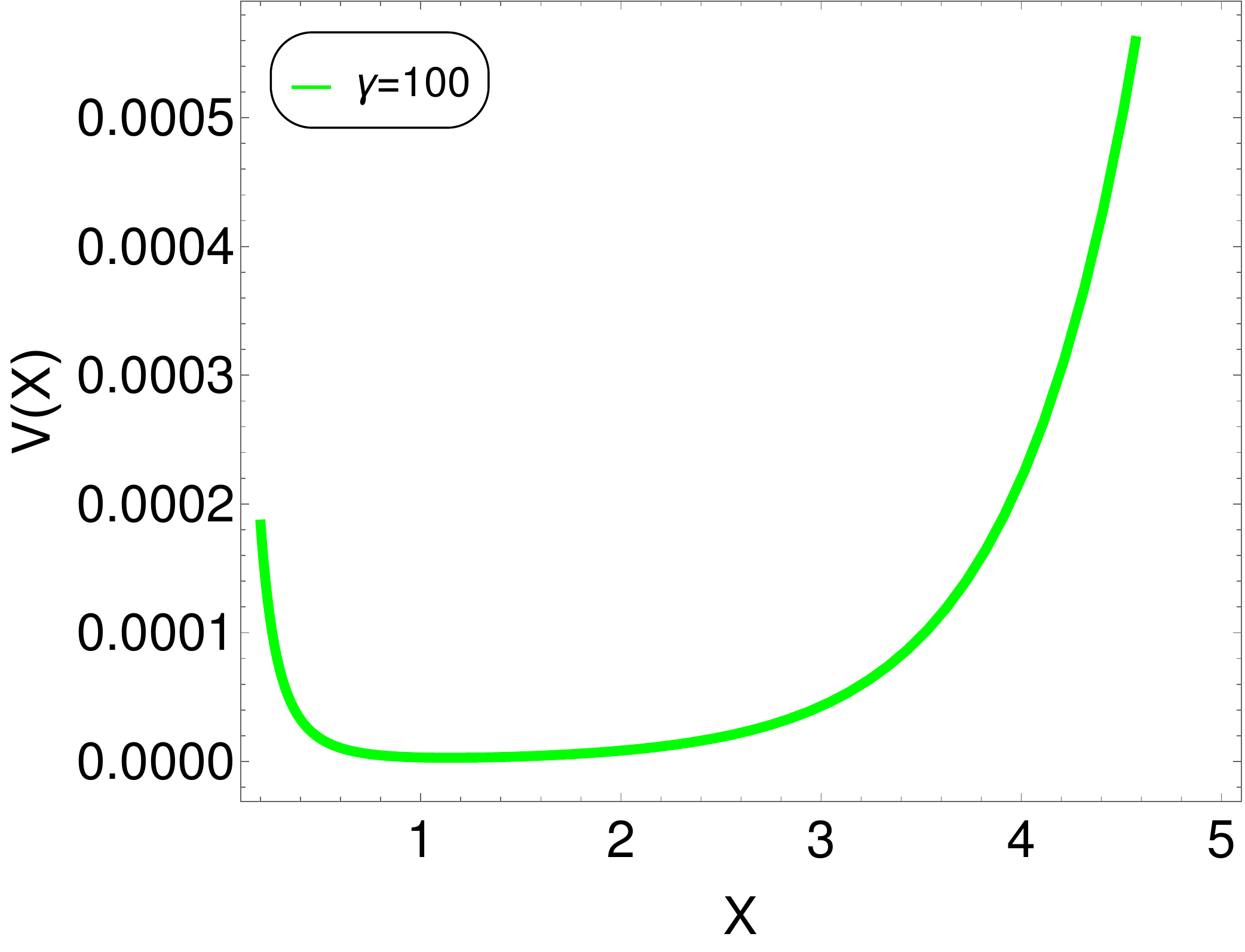}
  	\caption[]{\sl \small{The left panel shows the shape of the scalar potential for small field approximation. On the right panel the shape of the potential in the large field approximation is presented. Here we use $\lambda/\mu^2=1.76$, $\mu=0.08$, $M=1$ and $u=1/2$.}}
  	\label{Vapprox}
  \end{figure}

  Next we consider a small field approximation where,
  \begin{eqnarray}
  \varPhi\ll\frac{1}{\sqrt{\gamma}}
  \end{eqnarray}
with $\gamma<1$. In this scenario  we find that,
  \begin{eqnarray}
  \dfrac{dX}{d\varPhi}=\dfrac{1}{\sqrt{2u}}
  \end{eqnarray}
  which results to
  \begin{eqnarray}
  \varPhi=\sqrt{2 u}X\; .
  \end{eqnarray}
  Then  the effective potential ~(\ref{newff}) in this case receives the form,
  \begin{equation}\label{Vsfa}
  V\left(X\right) =	u\mu^2X^2+\frac{\lambda\left(1-2uX^2\right)^2}{4\left(1-2\gamma u X^2\right)^2}~\cdot 
  \end{equation}
The shape of the potential is shown in Figure ~\ref{Vapprox}    for both cases. In the small field approximation  the factor $\gamma X^{2}$ plays an important role on the predictions of the model. If the contribution of $\gamma$ in the potential \ref{Vsfa} is very small ($\gamma X^{2}\ll{1}$) then the tensor-to-scalar ratio is outside the Planck $2\sigma$ bounds. However, as  $\gamma X^{2}$ increased the denominator in \ref{Vsfa} becomes important and some solutions (depending upon the other parameters) consistent with the CMB observables appear.

Next,  we will compute the probability of the inflaton tunneling  from the false vacuum at $X_-$ to the true one located at  $X=0$. 
Following Coleman and deLucia work~\cite{Coleman:1980aw}, the analysis will take place for the case where the field $X$ interacts with  gravity.
Then, the Euclidean action of the coupled field $X$ takes the form
\begin{equation}
	S_E=\int d^4x \sqrt{g}\left(-\frac 1{2} R + \frac 12 \partial_{\mu}X \partial^{\mu}X + V(X)\right). \label{actionE}
\end{equation}
 The presence of gravity has important  implications to the decay of the false vacuum.   Moreover, note also that in the context of general relativity 
 there is an absolute minimum associated with the the positive cosmological constant (de Sitter spacetime), in contrast to the field theory case where only 
 potential and energy differences matter.

 Denoting with $S_E$  the Euclidean  action, the tunneling probability is defined as a decay rate per unit volume per time by 
  \begin{equation}
  \Gamma=Ae^{-B}, \qquad {\rm with} \qquad  B=S_E(X)-S_E(X_-)\; . \label{tunnelrate}
  \end{equation}
As in reference~\cite{Coleman:1980aw}, we look for $O(4)$ symmetric  solutions in a Euclidean four-space which  is  described by the  metric
  \begin{equation} 
  ds^2=d\rho^2+\sigma(\rho)^2(d\Omega_3)^2, \label{metricE}
  \end{equation}
  where $(d\Omega_3)^2$ is the metric of the unit 3-sphere. 
  The curvature of the 3-sphere at a given $\rho$ in~\eqref{metricE} is given by the scale factor $\sigma(\rho)$. 
   The  field equations for the scalar field $X$ and the scale factor $\sigma$ in the time-like case are 
  \begin{align}
  &X''+3\frac{\sigma'}{\sigma}X'=\frac{\partial V}{\partial X}, \label{phiE}\\
  & \sigma'^2=1+\frac{1}{3} \sigma^2\left(\frac{1}2 X'^2-V(X)\right)~, \label{scaleE} 
  \end{align}
  where the  derivatives $X',\sigma'$ etc  are taken with respect to the `time'-variable $\rho$. For the system of equations \eqref{phiE} and \eqref{scaleE}, we adopt the simplest solutions~\cite{Antoniadis:2020stf}, which for $\rho \in [0, H_{\pm}^{-1}\, \pi]$ are given by
  \begin{equation}
  X(\rho)=X_{\pm}, \quad \sigma(\rho)=\frac{1}{H_{\pm}} \sin(H_{\pm} \rho),  \label{dSsol}
  \end{equation}
where from the field equation $\frac 12 {H'}^2=3 H^2-3 V(X)$ at the minima $H'_{\pm}=0$, 
it turns out that
$ H_{\pm}=\sqrt{\frac{V(X_{\pm})}{3}}$.  Using  the solution \eqref{dSsol}  the action is found to be~\cite{Antoniadis:2020stf}
  \begin{equation}
  S_E(X_{\pm})=-2\pi^2\int d\rho\frac{1}{H_{\pm}^3}\sin^3(H_{\pm}\rho) V(X_{\pm})=-\frac{24\pi^2}{V(X_{\pm})}. \label{actionsdS}
  \end{equation}
  \begin{equation}
  H_c^2=-\frac{V_{XX}(X_+)}{4}-\frac{\Delta V}{3}, \label{H_c}
  \end{equation}
  where $V_{XX}= {\partial^2 V}/{\partial X ^2}$ and as above, $\Delta V=V(X_+)-V(X_-)$ is the height of the barrier. The  tunneling coefficient $B$ introduced in \eqref{tunnelrate} is then computed from \eqref{actionsdS} and reads
  \begin{equation}
  B=S_E(X_+)-S_E(X_-)=-\frac{24\pi^2}{V(X_+)}+\frac{24\pi^2}{V(X_-)}\; .
  \end{equation}

In order to estimate the parameter $B$ involved in the decay rate it is sufficient to compute  the potential at $X_{\pm}$. This follows in the next subsection.

  \subsection{Numerical Results}
   The post-inflationary Universe is described by radiation, matter and vaccum energy density. Furthermore, it assumed that the bubble nucleation rate $\Gamma$ in the past satisfies  $\Gamma\geq \Gamma_{0}$ where  $\Gamma_{0}$ is its 
  current Minkowski space value. The constraint on the nucleation rate $\Gamma_{0}$ from the post-inflationary era is given in~\cite{Markkanen:2018pdo}, which further implies the bound, $B_{0}\gtrsim 540$.
  Before presenting the  numerical details, we first  mention the scanning ranges. The shape of the potential and the existence of false minima depend upon two parameters, the  ratio $\lambda/\mu^2$  and $\gamma$. From the previous analytical calculation we have already found that false minima occur within the following bounds:
  \begin{equation}\label{bounds2}
  1<\frac{\lambda}{\mu^2}\leq\frac{27}{16}, \quad  \dfrac{(1-\sqrt{\frac{11}{27}})}{2}\leq\gamma<1~. 
  \end{equation} 
In the following the value of $ \mu$ will be fixed to  $ \mu=\frac 12$. Hence, we will perform the numerical scanning  varying $\lambda/\mu^2$  and $\gamma$ according to~(\ref{bounds2}). \\

There are two types of solutions that contribute to vacuum decay in de Sitter space: the   CdL solution~\cite{Coleman:1980aw} which crosses the barrier through tunneling, and  the Hawking-Moss solution(HM)~\cite{Hawking:1981fz}  where the inflaton is on the top of the barrier (for a more detailed discussion see~\cite{Antoniadis:2020stf}).  In ~(\ref{dSsol}) we defined $H_{-}$ which is actually the background Hubble rate in the false  vacuum. The condition for HM solutions to contribute to vacuum decay is $H \to H_c$, where $H_{c}$ is the critical Hubble rate defined in equation ~(\ref{H_c}). The second term  of the  critical Hubble rate generally contributes significantly if  the height  difference  between the top of the barrier and the false vacuum is comparable to the Planck mass. 

\begin{figure}	
	\begin{subfigure}{.5\textwidth}
		\centering
			\includegraphics[width=1\linewidth]{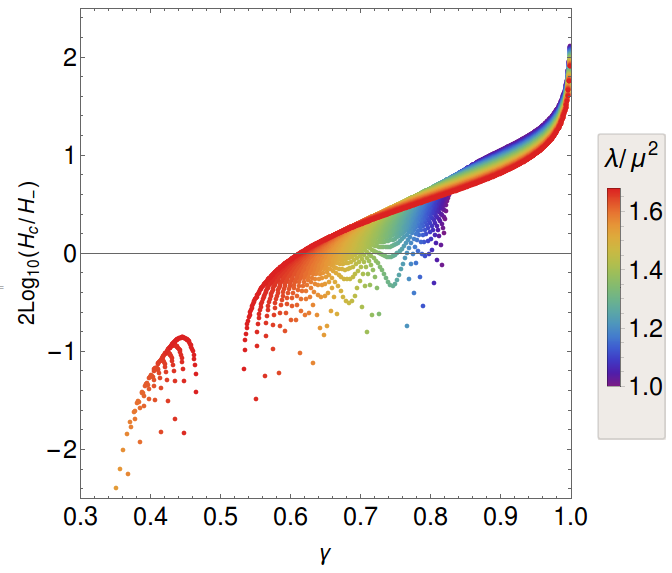}
	\end{subfigure}
	\begin{subfigure}{.5\textwidth}
		\centering
	 	\includegraphics[width=1\linewidth]{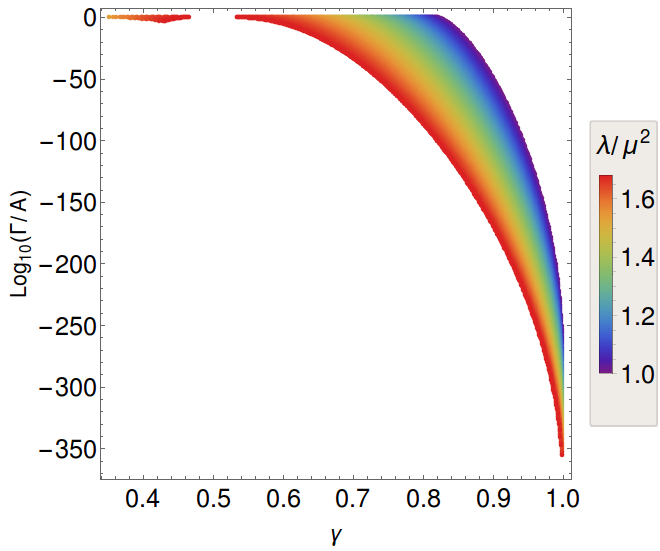}
		\end{subfigure}
	\begin{subfigure}{.5\textwidth}
		\centering
		\includegraphics[width=1\linewidth]{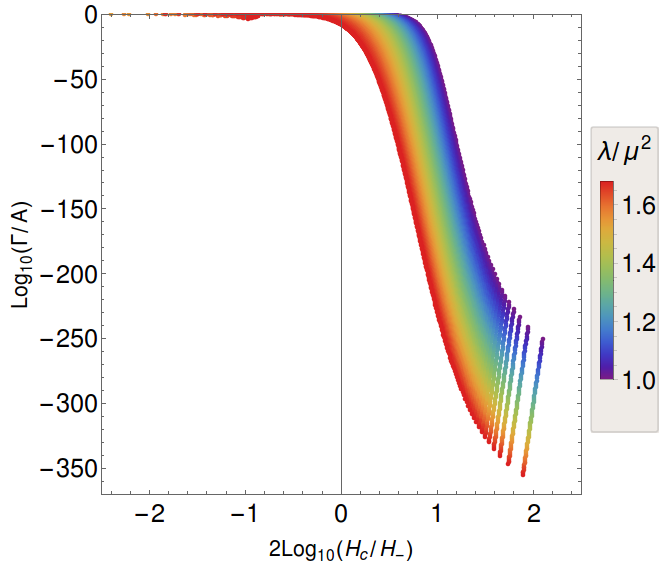}
	\end{subfigure}%
	\begin{subfigure}{.5\textwidth}
		\centering
			\includegraphics[width=1\linewidth]{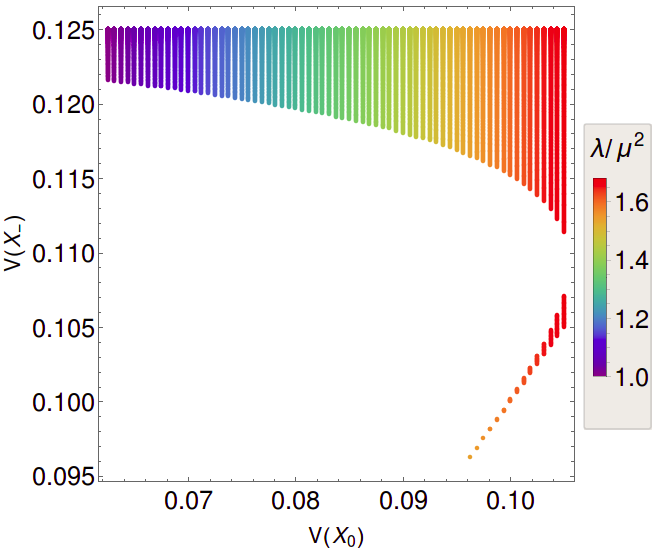}
		\end{subfigure}%
		\begin{center}
			\begin{subfigure}{.5\textwidth}
				\centering
				\includegraphics[width=1\linewidth]{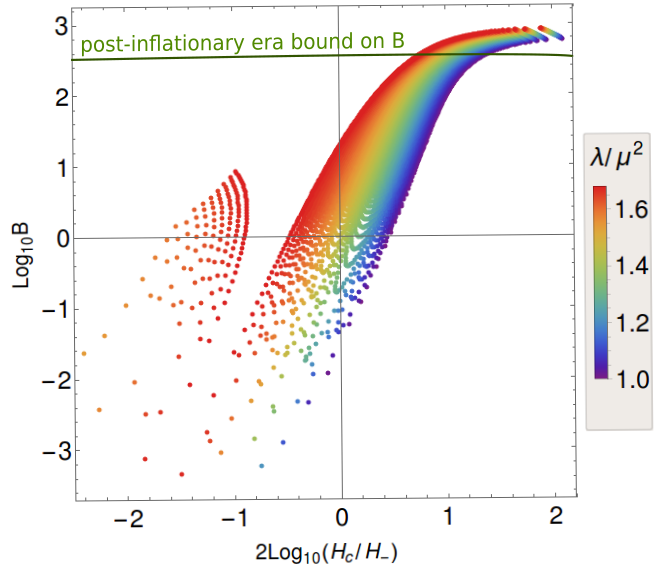}	
			\end{subfigure}
		\end{center}
	\caption[]{\sl\small{Figure shows the different results for the probability that the inflaton tunnels from the false vacuum at $X_-$ to the true vacuum $X=0$ . Here we use $\mu=0.5$.}}
	\label{cdl}
\end{figure}
 
  In Figure \ref{cdl}, the top plot  of the left panel  shows  $\gamma$ vs the ratio of the Hubble scale and the upper plot of the right panel shows  $\gamma$ vs the decay width  where the \emph{rainbow} color bar shows how these solutions vary with the ratio of $\lambda/\mu^2$. We see from the left panel upper plot that for low Hubble rate we have solutions corresponding  to HM instanton which contribute to vacuum decay. Recall that the HM instanton solution is the one for which the inflaton stays at the top of the barrier $X(\rho)=X_+$. This  describes the situation where  the inflaton climbs up the potential barrier instead of tunneling. On the other hand  when the Hubble rate increases i.e. $H_-<H_c$, standard CdL solutions appear. On the top left panel of Figure \ref{cdl},  the  solutions below the horizontal black line correspond to HM. On the other hand, those solutions which are placed  above the black horizontal line correspond to the CdL case. The middle-left panel of Figure \ref{cdl} shows the plot between the ratio of Hubble scales and the decay rate while on the middle-right panel one can see  the value of the potential at both minima.  In these scans we only show plots with solutions in which $V(X_{-})>V(X_{0})$, so the motion of the inflaton field is from right to left.  The lower panel shows the plot between the ratio of Hubble scales and the parameter $B$ where as horizontal line in the figure shows the bound on $B$
  coming from the post-inflationary era. Based on the numerical value of $B$  we can categorize our false vacuum into stable, meta-stable and unstable. From the plot we see that $B >B_{0}$ (where $B_{0}=540$) is a region where $\Gamma<\Gamma_{0}$ and we can infer that false vacuum is stable. Similarly  $B\simeq B_{0}$ corresponds to meta-stable and $B<B_{0}$ represents the unstable region.

\section{Conclusion}

	In this work we have investigated various cosmological implications of a generic model based on the extension of the SM gauge 
	symmetry by a $U(1)_{B-L}$  factor, laying special emphasis on the issues of inflation, leptogenesis and baryogenesis as well as the physics  of gravitino. The model can be naturally embedded in a unified  gauge group with symmetry breaking scale around $M_{GUT}=2\times 10^{16}$ GeV.
The spectrum of the model consists of the MSSM content
extended by a neutral singlet field $S$,  and a pair of 
Higgs MSSM singlets $(H,\bar H)$   carrying opposite charge under $U(1)_{B-L}$. An appropriate  R-symmetry prevents all the dangerous terms in the superpotential whereas  a suitable fourth order non-renormalizable term providing
Majorana mass to the right-handed  neutrino is left intact to realise the seesaw mechanism. 
The most general effective potential consists of the F- and D-parts as well  and contributions  coming from soft supersymmetry breaking terms. 
The F-part arises from the K\"ahler fucntion, assuming standard no-scale supergravity and the 
D-terms contain the usual contributions associated with the gauge sector.
Under mild constraints on the parameters involved in the scalar potential, it is readily realised that the inflationary scenario is naturally implemented.
The cosmological observables, including the tensor-to-scalar ratio $r$, 
the spectral index $n_s$ etc,  are computed  and discussed in  detail 
for  various limiting cases of the effective potential.
In the  present analysis, a wide  viewpoint  is taken, 
 to encompass future  perspectives on  the possible determination of the supersymmetry breaking scale, $m_{SUSY}$. 
 Thus, in this context, varying $m_{SUSY}$ from a few TeV up to $10^6$ TeV, 
   while fixing the spectral index to its central  value $n_s=0.9655$, we find that
 the tensor-to-scalar ratio  lies in the rangle   $r\in [10^{-2}-10^{-3}]$
 which is consistent with the latest Planck data. 

We examine in detail the physics related to the nature of gravitino, 
considering  possible scenarios, including that it is the lightest 
supersymmetric particle (LSP). Among other possibilities,  it is found  that 
a stable LSP gravitino  is easily accommodated in our setup and as such,
it can be considered as a DM candidate. If the gravitino is not the LSP,
it is found  that a short lived gravitino is always viable as long as its mass is  $m_{3/2} >25$ TeV whereas the mass of  a long-lived one  should
lie in the region $10 < m_{3/2} <25$ TeV.

 For   high SUSY scales the reheating temperature $T_r$ is bounded by $\Omega_{LSP}$ whereas for a TeV scale SUSY case the reheating temperature is bounded by BBN predictions.
Furthermore the model predicts the inflaton mass to be around $10^{12}$ GeV. On the other hand, the   variation of the right-handed heavy Majorana
scale  depends on $T_r$  and ranges between $M_N\sim 10^{8}$ and $10^{12}$ GeV. Values close to the upper scale are sufficient to realise the see-saw mechanism and obtain
light neutrino masses consistent with the oscillation experiments. 
 Finally we identify regions of the parameter space where the 
 potential in addition to the true vacuum it also displays a false minimum. 
 For this particular case, we discuss quantum tunneling effects  with the
 inflaton field penetrating   the barrier,  and compute
 its decay width  to the true vacuum using standard 
 techniques developed by Coleman and deLucia~\cite{Coleman:1980aw}.

\section{Acknowledgements} 
The authors are thankful to Mansoor Ur Rehman and  Fariha K. Vardag for fruitful discussions. The work of GKL was supported by the ``Hellenic Foundation for Research and Innovation (H.F.R.I.) under the ``First Call for H.F.R.I. Research Projects to support Faculty members and
Researchers and the procurement of high-cost research equipment
grant'' (Project Number: 2251)''.

\end{document}